\documentclass[prd,reprint, nofootinbib,amsmath,amssymb,amsfonts,aps,showpacs,
superscriptaddress,eqsecnum, floatfix, longbibliography]{revtex4-2}
\usepackage[colorlinks=true,allcolors=blue]{hyperref}
\usepackage{enumerate}
\usepackage{mathrsfs}

\def\be{\begin{equation}}
\def\ee{\end{equation}}

\usepackage{array,multirow}
\newcolumntype{L}[1]{>{\raggedright\let\newline\\\arraybackslash\hspace{0pt}}m{#1}}
\newcolumntype{C}[1]{>{\centering\let\newline\\\arraybackslash\hspace{0pt}}m{#1}}
\newcolumntype{R}[1]{>{\raggedleft\let\newline\\\arraybackslash\hspace{0pt}}m{#1}}

\begin{document}

\title{Conserved charges of the extended Bondi-Metzner-Sachs algebra}
\author{\'Eanna \'E. Flanagan}%
 \email{eef3@cornell.edu}
\affiliation{ Department of Physics, Cornell University, Ithaca, New York 14853, USA }%

\author{David A.\ Nichols}
\email{dnichols@astro.cornell.edu}
\affiliation{Cornell Center for Astrophysics and Planetary Science (CCAPS),
Cornell University, Ithaca, New York 14853, USA}%

%
%
\newcount\hh
\newcount\mm
\mm=\time
\hh=\time
\divide\hh by 60
\divide\mm by 60
\multiply\mm by 60
\mm=-\mm
\advance\mm by \time
\def\hhmm{\number\hh:\ifnum\mm<10{}0\fi\number\mm}



\begin{abstract}
Isolated objects in asymptotically flat spacetimes in general
relativity are characterized by their conserved charges associated with the
Bondi-Metzner-Sachs (BMS) group.  These charges include total energy, linear momentum,
intrinsic angular momentum and center-of-mass location,
and, in addition, an infinite number
of supermomentum charges associated with supertranslations.
Recently, it has been suggested that the BMS symmetry algebra should be
enlarged to include an infinite number of additional symmetries known as superrotations.
We show that the corresponding charges are finite and well defined,
and can be divided into electric parity ``super center-of-mass'' charges and magnetic parity ``superspin'' charges.

The supermomentum charges are associated with ordinary gravitational-wave
memory, and the super center-of-mass charges are associated with
total (ordinary plus null) gravitational-wave memory, in
the terminology of Bieri and Garfinkle.  Superspin charges are associated with the ordinary piece of spin memory.
Some of these charges can give rise to black-hole hair, as described
by Strominger and Zhiboedov.  We clarify how this hair evades the no-hair
theorems.

\end{abstract}

\maketitle

\begin{widetext}
\tableofcontents
\end{widetext}

\section{\label{intro} Introduction}

Spacetimes which are asymptotically flat at future null infinity in
general relativity have a group of asymptotic symmetries known as the
BMS group \cite{1962RSPSA.269...21B,
1962RSPSA.270..103S,1962PhRv..128.2851S}.  Associated with each
generator ${\vec \xi}$ of this group and each cross section of future
null infinity, there is a conserved charge\footnote{By conserved charge we mean a charge that would be conserved in the absence of fluxes of radiation to null infinity.} $Q$ \cite{1981RSPSA.376..585A,1984CQGra...1...15D,
Wald:1999wa}.  These charges include all the charges associated with
the Poincar\'e group and, in addition, an infinite number of new
supermomentum charges associated with supertranslations.
The generators ${\vec \xi}$ of the BMS group are smooth vector fields on
future null infinity ${\mathscr I^+}$.

Recently Banks \cite{Banks:2003vp} and  Barnich and Troessaert
\cite{Barnich:2009se,Barnich:2011ct,2011JHEP...12..105B} have suggested
that a larger symmetry algebra might be physically relevant.
In particular, they suggested including vector fields called ``superrotations''
which contain
analytic singularities as functions on the conformal 2-sphere at
infinity.  These are formally infinitesimal symmetries of the theory,
but they cannot be exponentiated to yield smooth finite diffeomorphisms,
unlike the generators of the standard BMS group.
There is not yet a general theory for understanding when singular
vector fields of this type can be used to construct conserved charges
and fluxes at future null infinity, and there has been some debate in
the literature on the physical relevance or utility of the extended algebra.

One approach to this question is to determine whether the new
symmetries give rise to relations between $S$-matrix elements, like
the standard symmetries do \cite{Strominger:2013jfa,He:2014laa}.
This was shown to be the case for the tree level $S$-matrix by Kapec,
Lysov, Pasterski, and Strominger \cite{Kapec:2014opa}.

Another approach is to determine whether
the new symmetries give rise to well-defined and finite
classical conserved quantities.  Here we adopt this approach,
following Barnich and Troessaert \cite{2011JHEP...12..105B}
who showed that the charges associated with superrotations vanish in
the Kerr spacetime.
We show that the superrotation charges are in general finite.
There are two pieces of these charges, an electric parity piece and a
magnetic parity piece.  We call the electric parity charges
{\it super center-of-mass} charges, since they
are generalizations of the center-of-mass piece of special-relativistic
angular momentum.
We call the magnetic parity charges {\it superspin} charges,
since they are generalizations of the intrinsic angular momentum piece of special-relativistic
angular momentum.
In addition,  we show that
supermomentum charges are associated with ordinary gravitational-wave
memory and super center-of-mass charges are associated with total
(ordinary plus null) gravitational-wave memory, in the terminology of
Bieri and Garfinkle \cite{Bieri:2013ada}.
The superspin charges are associated the ordinary piece of ``spin
memory'', the new type of gravitational-wave memory discovered by
Pasterski, Strominger, and Zhiboedov \cite{Pasterski:2015tva}.

The paper is organized as follows.  Section \ref{sec:bms} reviews the
standard BMS group and algebra and also the extended BMS algebra.
In Sec.\ \ref{sec:charges}, we compute the conserved charges.
In Sec.\ \ref{sec:interpretation}, we make some remarks about the physical
significance of BMS and extended BMS charges and how they can be measured.

\section{\label{sec:bms} The BMS symmetry group and the extended BMS algebra}

\subsection{Asymptotically flat spacetimes in retarded Bondi coordinates}

We start by reviewing the definition of the BMS
symmetry group and its action on solutions of Einstein's equations
near future null infinity.  We closely follow the exposition of
Barnich and Troessaert \cite{Barnich:2009se,Barnich:2010eb}
as simplified and specialized by Strominger and collaborators
\cite{Strominger:2013jfa,He:2014laa,Kapec:2014opa,Strominger:2014pwa,Pasterski:2015tva}.
Our notation, however, will follow Barnich and Troessaert in maintaining
covariance with respect to the 2-sphere coordinates instead of using
the complex-coordinate convention used in Refs.\ \cite{Strominger:2013jfa,He:2014laa,Kapec:2014opa,Strominger:2014pwa,Pasterski:2015tva}.

Following Refs.\
\cite{1962RSPSA.269...21B,Barnich:2009se,Barnich:2010eb,Strominger:2013jfa,He:2014laa,Kapec:2014opa,Strominger:2014pwa,Pasterski:2015tva}
we use retarded Bondi coordinates $(u,r,\theta^1,\theta^2)$ near
future null infinity.  The metric has the form
\begin{eqnarray}
ds^2 &=& - U e^{2 \beta} du^2 - 2 e^{2 \beta} du dr \nonumber \\
&&+ r^2 \gamma_{AB} (d\theta^A -
{\cal U}^A du) (d\theta^B - {\cal U}^B du),
\label{metric}
\end{eqnarray}
where $A,B = 1,2$, and $U$, $\beta$, ${\cal U}^A$, and $\gamma_{AB}$
are functions of $u$, $r$, and $\theta^A$.  The four gauge conditions that are
imposed are $g_{rr} =0$, $g_{rA} =0$, and\footnote{One could further specialize the gauge by imposing $\partial_u {\rm det}(\gamma_{AB}) =0$, but in the context of the expansion in powers of $1/r$ this condition follows from Eqs.\ (\ref{gaugec}) and (\ref{metricfns}).}
\be
\partial_r {\rm det}(\gamma_{AB}) =0.
\label{gaugec}
\ee

We now expand the metric functions as series in $1/r$.
The order in $1/r$ at which the various expansions start can be
deduced from the covariant definition of asymptotic flatness at future null
infinity \cite{1984ucp..book.....W}.
The expansions are\footnote{Some of the higher-order terms in these expansions are not needed in this section but will be needed in Sec.\ \ref{sec:stationary} below.} \cite{1962RSPSA.269...21B,Barnich:2009se,Barnich:2010eb,Strominger:2013jfa,He:2014laa,Kapec:2014opa,Strominger:2014pwa,Pasterski:2015tva}:
\begin{subequations}
\label{metricfns}
\begin{eqnarray}
\beta &=& \frac{\beta_0}{r} + \frac{\beta_1}{r^2} + \frac{\beta_2}{r^3}+O(r^{-4}), \\
\label{Uexpand}
U &=& 1 - \frac{2 m}{r} - \frac{2 {\cal M}}{r^2} +O(r^{-3}), \\
\gamma_{AB} &=& h_{AB} + \frac{1}{r} C_{AB} + \frac{1}{r^2} D_{AB} +
\frac{1}{r^3} E_{AB} \\ \nonumber
&& +
O(r^{-4}),\\
{\cal U}^A &=& \frac{1}{r^2} U^A + \frac{1}{r^3} \bigg[ - \frac{2}{3} N^A
+ \frac{1}{16} D^A(C_{BC} C^{BC})
\nonumber \\
&& + \frac{1}{2} C^{AB} D^C C_{BC} \bigg]
+ O(r^{-4}).
\label{NAdef}
\end{eqnarray}
\end{subequations}
Here the various coefficients on the right-hand sides are functions of
$(u,\theta^A)$ only.
The metric $h_{AB}(\theta^C)$ is the fixed round metric on the unit 2-sphere.
In adapted coordinates $(\theta,\varphi)$, it is
$d\theta^2 + \sin^2 \theta d\varphi^2$, but we will use general coordinates
$\theta^A$ and retain two-dimensional covariance throughout.
We adopt the convention that capital Roman indices (e.g., $A$, $B$) are raised and lowered
with $h_{AB}$, and we denote by $D_A$ the covariant derivative associated with $h_{AB}$.
There are three important, leading-order functions in the metric's expansion coefficients
\cite{1962RSPSA.269...21B,Barnich:2009se,Barnich:2010eb,Strominger:2013jfa,He:2014laa,Kapec:2014opa,Strominger:2014pwa,Pasterski:2015tva}:
the Bondi mass aspect $m(u,\theta^A)$, the angular-momentum
aspect\footnote{Our definition of the angular-momentum aspect $N_A$ [cf.\
Eq.\ (\ref{NAdef})] follows
Ref.\ \cite{Pasterski:2015tva} rather than the one used in
Ref.\ \cite{Barnich:2010eb}.
Our definition also coincides with that used in
Sec.\ 5.6 of the book by Chrusciel, Jezierski, and Kijowski \cite{2002hftr.book.....C}, up to a factor of $-3$.}
$N^A(u,\theta^A)$, and the symmetric tensor
$C_{AB}(u,\theta^A)$ whose derivative
\be
N_{AB} = \partial_u C_{AB}
\label{news}
\ee
is the Bondi news tensor.

Imposing the gauge condition (\ref{gaugec}) now yields the constraints
\begin{subequations}
\label{con1}
\begin{eqnarray}
h^{AB} C_{AB} &=& 0, \\
D_{AB} &=& C_{CD} C^{CD} h_{AB}/4 +
{\cal D}_{AB}, \\
E_{AB} &=& C_{CD} {\cal D}^{CD} h_{AB}/2 + {\cal E}_{AB},
\end{eqnarray}
\end{subequations}
where the tensors ${\cal D}_{AB}$ and ${\cal E}_{AB}$ are traceless.

We assume the following behavior
of the stress-energy tensor as $r \to \infty$:
\begin{subequations}
\label{Tabscaling}
\begin{align}
T_{uu} =& \frac{1}{r^2} {\hat T}_{uu}(u,\theta^A) + O(r^{-3}), \\
T_{rr} =& \frac{1}{r^4} {\hat T}_{rr}(u,\theta^A)
+ \frac{1}{r^5} {\tilde T}_{rr}(u,\theta^A)
+ O(r^{-6}), \ \ \\
T_{uA} =& \frac{1}{r^2} {\hat T}_{uA}(u,\theta^A) + O(r^{-3}), \\
T_{rA} =& \frac 1{r^3} {\hat T}_{rA}(u,\theta^A)  + O(r^{-4}) \, ,
\\ T_{AB} =& \frac{1}{r} {\hat T}(u,\theta^A) h_{AB}+ O(r^{-2})\, ,
\end{align}
\end{subequations}
together with $T_{ur} = O(r^{-4})$.
These assumptions are motivated by the behavior of radiative scalar-field
solutions in Minkowski spacetime\footnote{The case considered in
  Refs.\
  \cite{Strominger:2013jfa,He:2014laa,Kapec:2014opa,Strominger:2014pwa,Pasterski:2015tva}
corresponds to ${\hat T}=0$, which applies for example to conformally invariant fields.}.
Imposing stress-energy conservation yields
\begin{subequations}
\label{ccc}
\begin{eqnarray}
\partial_u {\hat T}_{rA} &=&D_A {\hat T}, \\
\partial_u {\hat T}_{rr} &=& - 2 {\hat T},
\end{eqnarray}
\end{subequations}
from $O(A,5)$ and $O(r,3)$, respectively, where
$O(\alpha,n)$ means the $O(r^{-n})$ piece of the $\alpha$ component of
$\nabla^\beta T_{\alpha\beta} =0$.  Combining Eqs.\ (\ref{ccc}) we
can write
\be
{\hat T}_{rA}(u,\theta^A) = {\check T}_{rA}(\theta^A) - \frac 12
D_A {\hat T}_{rr}(u,\theta^A),
\ee
for some function ${\check T}_{rA}(\theta^A)$.

We now impose Einstein's equations $G_{ab} = 8 \pi T_{ab}$, in units with
$G = 1$.  We adopt the shorthand notation that $O(\alpha\beta,n)$ means
the $O(r^{-n})$ piece of the $(\alpha\beta)$ component of Einstein's
equations.
We obtain
\begin{subequations}
  \label{con2}
\begin{eqnarray}
U_A &=& - D^B C_{AB}/2, \\
\beta_0 &=& 0, \\
\beta_1 &=& - \frac{1}{32} C_{AB} C^{AB} - \pi {\hat T}_{rr},
\\
\beta_2 &=& - \frac{1}{12} C_{AB} D^{AB} - \frac{2 \pi}{3} {\tilde T}_{rr}, \\
\label{calDeqn}
D^A {\cal D}_{AB} &=& - 8 \pi {\check T}_{rB}, \label{29e}\\
\partial_u {\cal D}_{AB} &=& 0,
\end{eqnarray}
\end{subequations}
from $O(rA,2)$, $O(rr,3)$, $O(rr,4)$, $O(rr,5)$, $O(rA,3)$, and $O(AB,1)$,
respectively.  Note that it follows from Eq.\ (\ref{calDeqn}) that
\be
{\cal D}_{AB} =0
\label{calD0}
\ee
in vacuum.
We also obtain from $O(uu,2)$ and $O(uA,2)$
evolution equations for the
Bondi mass aspect $m$ and the angular-momentum aspect $N_A$
\cite{Barnich:2010eb,Strominger:2013jfa,Pasterski:2015tva}:
\begin{subequations}
\label{dots}
\begin{eqnarray}
\label{dotm}
{\dot m} &=& - 4 \pi {\hat T}_{uu} - \frac{1}{8} N_{AB} N^{AB}
+ \frac{1}{4} D_A D_B N^{AB}, \ \ \ \\
\label{dotNA}
{\dot N}_A& = & - 8 \pi {\hat T}_{uA} + \pi D_A \partial_u {\hat T}_{rr}
 + D_Am + \frac{1}{4}
D_B D_A D_C C^{BC} \nonumber \\
&& - \frac{1}{4} D_B D^B D^C C_{CA} + \frac{1}{4} D_B (N^{BC} C_{CA})
\nonumber \\
&&+ \frac{1}{2} D_B N^{BC} C_{CA}.
\end{eqnarray}
\end{subequations}
Here dots denote derivatives with respect to $u$.
The leading-order components of the Weyl tensor for these solutions are listed in Appendix \ref{app:weyl}.

\subsection{\label{sec:bmsgroup} BMS symmetry group}

BMS symmetries are diffeomorphisms of future null infinity, ${\mathscr
  I^+}$,
 to itself that preserve its intrinsic geometric properties
\cite{bobg,1984ucp..book.....W}.
Explicitly, in Bondi coordinates, the diffeomorphism $\psi$ takes the point
$(u,\theta^A)$ on ${\mathscr I^+}$ to $({\bar u},{\bar \theta}^A)$, where
\begin{subequations}
\begin{eqnarray}
{\bar u} &=& \frac{1}{w(\theta^A)} \left[ u + \alpha(\theta^A)
\right], \\
{\bar \theta}^A &=& {\bar \theta}^A(\theta^B).
\end{eqnarray}
\label{bmsfinite}
\end{subequations}
Here the mapping $\theta^A \to {\bar \theta}^A(\theta^B)$ must be a
conformal isometry of the 2-sphere, of which there is a six-parameter
group, and the corresponding function $\omega$ is defined
by $\psi_* h_{AB} = \omega^{-2} h_{AB}$.  The function $\alpha$ can be
freely chosen.  The corresponding infinitesimal symmetries are
${\bar u} = u + \xi^u$, ${\bar \theta}^A = \theta^A + \xi^A$,
where the vector field ${\vec \xi}$ on ${\mathscr I^+}$ is
\begin{align}
{\vec \xi} &={} \xi^u \partial_u + \xi^A \partial_A \nonumber \\
&={} \left[ \alpha(\theta^A) + \frac{1}{2} u D_A Y^A(\theta^B)
\right] \partial_u + Y^A(\theta_B) \partial_A.
\label{bmsinfinitesimal}
\end{align}
Here $Y^A(\theta^B)$ must be a conformal Killing vector on the
2-sphere---i.e., be a solution of
\be
2 D_{(A} Y_{B)} - D_C Y^C h_{AB}=0.
\label{CKV}
\ee
The general solution can be written as
\be
Y^A = D^A \chi + \epsilon^{AB} D_B \kappa,
\label{lorentz}
\ee
where $\chi$ and $\kappa$ are $l=1$ spherical harmonics, that is,
solutions of $(D^2 + 2) \chi =0$ and $(D^2 + 2) \kappa =0$, where $D^2
= D_A D^A$.  These solutions comprise the Lorentz algebra,
with the three electric parity solutions $D^A \chi$ corresponding to
boosts, while the three magnetic parity solutions $\epsilon^{AB} D_B
\kappa$ correspond to rotations.

The symmetry vector fields ${\vec \xi}$ can be extended from future
null infinity ${\mathscr I^+}$ into the interior of the spacetime to give
approximate asymptotic Killing vectors by demanding that they
maintain the retarded Bondi coordinate conditions and the assumed
scalings with $r$ of the metric components.  This gives
\cite{Barnich:2010eb,Strominger:2013jfa}
\begin{eqnarray}
{\vec \xi} &=& f \partial_u + \left[ Y^A - \frac{1}{r} D^A f +
  \frac{1}{2 r^2} C^{AB} D_Bf + O(r^{-3}) \right] \partial_A \nonumber
\\
&&- \bigg[  \frac{1}{2} r D_A Y^A - \frac{1}{2} D^2 f
- \frac{1}{2 r} U^A D_Af \nonumber \\
&& + \frac{1}{4 r} D_A (D_B f C^{AB}) + O(r^{-2}) \bigg] \partial_r,
\label{generator}
\end{eqnarray}
where
\be
f(u,\theta^A) = \alpha(\theta^A) + \frac{1}{2} u D_B Y^B(\theta^A).
\label{fdef}
\ee
Under these transformations, the metric transforms via pullback as
$g_{ab} \to \psi_* g_{ab} = g_{ab} + {\cal L}_{\vec \xi} g_{ab}$.
This yields the following transformations of the metric functions \cite{Barnich:2010eb,Strominger:2013jfa},
\begin{widetext}
\begin{subequations}
\label{transforms}
\begin{eqnarray}
\delta m &=& f {\dot m} + \frac{1}{4} N^{AB} D_A D_B f
+ \frac{1}{2} D_A f D_B N^{AB}
+ \frac{3}{2} m \psi + Y^A D_A m
+ \frac{1}{8} C^{AB} D_A D_B \psi,  \\
\label{deltaCAB}
\delta C_{AB} &=& f N_{AB} - 2 D_A D_B f + h_{AB} D^2 f- \frac{1}{2} \psi C_{AB} + {\cal L}_{\vec Y} C_{AB}, \\
%
%
%
\delta N_A &=&  (f\partial_u +\mathcal L_{\vec Y} + \psi) N_A +
3m D_A f
- \frac 12 \mathcal D_{AB} D^B f + \pi (\partial_u \hat T_{rr}) D_A f 
- \frac\pi 2 \hat T_{rr} D_A \psi  \nonumber \\
&&  - \frac 34 D_B f(D^B D^C C_{CA} - D_A D_C C^{BC})
+ \frac 34 C_{AB} N^{BC} D_C f \, .
\label{NAtransform}
\end{eqnarray}
\end{subequations}
\end{widetext}
where on the right-hand sides $N_{AB}$ is the Bondi news tensor (\ref{news}),
overdots denote derivatives with respect to $u$, ${\cal L}$ is a Lie derivative,
$f$ is defined by Eq.\ (\ref{fdef}), and $\psi$ is defined by
\be
\psi \equiv D_A Y^A.
\ee

\subsection{Terminology for regions of future null infinity}

If the Bondi news tensor $N_{AB}$ vanishes in a region of future null
infinity in a given BMS frame, then it will vanish in that region in
all BMS frames.  The region is then called {\it nonradiative}.
We call a region of ${\mathscr I}^+$ {\it stationary} if the spacetime
is stationary in a neighborhood of that region.
If a region is stationary then it
must be nonradiative \cite{bobg}.  However the converse is not true,
for example, a linear superposition of the linearized gravitational
fields of two point particles with a relative boost is nonradiative
but nonstationary.

When computing charges later in this paper we will specialize, for
simplicity, to nonradiative regions of ${\mathscr I}^+$, and sometimes
in addition specialize further to stationary regions.
We will consider {\it nonradiative to nonradiative transitions},
that is, spacetimes which possess a nonradiative region of ${\mathscr
  I}^+$, followed by a radiative region, followed by another
nonradiative region.  We will also consider stationary to stationary
transitions.

\subsection{\label{sec:stationary} Canonical Bondi frame for
  stationary vacuum regions}

Consider a stationary region of ${\mathscr I^+}$ in which the
leading-order stress-energy components (\ref{Tabscaling}) as well as
the subleading components vanish.
Then, there exists a preferred, canonical Bondi frame in which the
metric
takes a simple form, as we now review.

First, it is known that the news tensor (\ref{news}) must vanish in stationary regions \cite{bobg}, so that $C_{AB}$ is independent of $u$.
Next, it follows from the evolution equation (\ref{dotm}) for the Bondi mass aspect that $m$ is also independent of $u$.
From the $O(ur,4)$ component of Einstein's equation we now obtain
\be
6 {\cal M} + D_A N^A + \frac{3}{16} C_{AB} C^{AB}
+  \frac{3}{4} D_A C^{AB} D^C C_{CB}
 = 0 \,,
\label{calM}
\ee
which will be useful below.

We now specialize to Bondi frames in which the angular momentum aspect is independent of $u$,
so that
\be
\partial_u N_A = 0.
\label{specialize}
\ee
The existence of such frames is established in Sec.\ 6.7 of Ref.\ \cite{2002hftr.book.....C}.
We will show that it is possible to further specialize the frame to the preferred, canonical one.

We first derive some properties of the Bondi mass aspect $m$ and
angular-momentum aspect $N_A$ under the above assumptions.
From the evolution equation (\ref{dotNA}) with $N_{AB}={\dot N}_A=0$, we
obtain $4 D_A m - \epsilon_{AB}D^B \gamma =0$, where
$\gamma = h^{BD} \epsilon^{AC} D_C D_B C_{AD}$.  It follows that
\be
m(\theta^A) = m_0,
\label{mconst}
\ee
a constant.
Taking next the subleading $O(uA,3)$ component and using Eq.\ (\ref{calD0})
as well as Eq.\ (\ref{calM}) to
eliminate the subleading mass function ${\cal M}$ yields
\be
D^2 N_A  + N_A= - 3 D^B (m C_{AB}) + w_A.
\label{NAeqn}
\ee
Here $w_A$ is an expression quadratic and cubic in $C_{AB}$ and its derivatives,
whose precise form will not be needed.
We have also assumed that the subleading angular momentum aspect is
independent of $u$.  Equation \eqref{NAeqn} will be
useful below.

We now derive the transformation to the preferred, canonical Bondi frame.
The tensor $C_{AB}$ can be decomposed into electric and magnetic parity
pieces,
\be
C_{AB} = (D_A D_B - \frac{1}{2} h_{AB} D^2) \Phi
+ \epsilon_{C(A} D_{B)} D^C \Psi,
\label{split}
\ee
where, without loss of generality, $\Phi$ and $\Psi$ have no $l=0,1$ components.
Taking the magnetic parity part of the time-evolution equation (\ref{dotNA}) for
the angular-momentum aspect by contracting it with $\epsilon^{AB}
D_B$, using $N_{AB}={\dot N}_A=0$, and commuting indices using
$R_{ABCD} = h_{AC} h_{BD} - h_{AD} h_{BC}$ gives
$ (D^2 + D^4/2) \Psi=0$.  This forces\footnote{\label{foot:ProjL01}
The inverse of the angular differential operator $D^2+ D^4/2$ on the space of
functions with no $l=0,1$ pieces
is given explicitly in Eq.\ (2.17) of Ref.\ \cite{Pasterski:2015tva}
and also in Appendix C of Ref.\ \cite{Bieri:2013ada}.}
the magnetic part $\Psi$ to vanish.
Next, by using a BMS transformation with $Y^A=0$ and $\alpha =
\Phi/2$, we see from Eq.\ (\ref{deltaCAB}) that we can also make the electric
part $\Phi$ vanish, so that $C_{AB}=0$.

Equation (\ref{NAeqn}) for the angular-momentum aspect now reduces to
\be
D^2 N_A  + N_A=0.
\label{NAeqn1}
\ee
We decompose $N_A$ into electric parity and magnetic parity
pieces,
\be
N_A = D_A \Upsilon + \epsilon_{AB} D^B \Theta,
\label{NAhomsol1}
\ee
where $D^2 \Upsilon = D_A N^A$ and $D^2 \Theta = -\epsilon^{AB} D_A N_B$.
Substituting into Eq.\ (\ref{NAeqn1}) now shows that $\Upsilon$ and
$\Theta$ satisfy
\be
(D^2 + 2) \Upsilon=0, \  \ \ \ \ (D^2 + 2) \Theta =0
\label{NAhomsol2}
\ee
(i.e., they are both $l=1$ spherical harmonics).  Thus, the solutions of Eq.\
(\ref{NAeqn1}) for the angular-momentum aspect coincide with the solutions
discussed after Eq.\ (\ref{CKV}) of the conformal Killing vector
equation: three electric parity conformal Killing vectors and
three magnetic parity Killing vectors.

We now perform a BMS transformation with $\alpha = -\Upsilon/(3 m_0)$ and
$Y^A=0$.
From the transformation law (\ref{deltaCAB}) for $C_{AB}$, we find that this
transformation does not alter the gauge specialization $C_{AB}=0$ that
we have already achieved, because the differential operator $2 D_A D_B -
h_{AB} D^2$ on the right-hand side annihilates $l=1$ spherical harmonics.
The effect of the transformation
is to set the electric parity piece of $N_A$ to zero, from Eq.\
(\ref{NAtransform}).
Since the electric parity piece of $N_A$ encodes information about the
center of mass, this transformation corresponds roughly to
translating to the center-of-mass
frame.  The remaining magnetic parity piece of $N_A$ encodes the intrinsic
angular momentum.

To summarize, we have achieved a Bondi frame in which
\begin{subequations}
\label{stat}
\begin{eqnarray}
\label{mstationary}
m(\theta^A) &=& m_0 = {\rm constant}, \\
\label{CABstationary}
C_{AB}(\theta^A) &=& 0, \\
N_A(\theta^A) &=& {\rm magnetic\ parity,\ }l=1.
\end{eqnarray}
\end{subequations}
We call the frame which satisfies these conditions the canonical frame.
The explicit construction of this frame in the Kerr spacetime can be found in Appendix C.7 of Ref.\ \cite{2002hftr.book.....C}.

\subsection{\label{sec:bmsextended} Extended BMS algebra}

The BMS algebra of approximate Killing vector fields ${\vec \xi}$
described above consists of vector fields that are smooth and finite
on future null infinity ${\mathscr I^+}$.  Relaxing this requirement,
Banks \cite{Banks:2003vp} and  Barnich and Troessaert
\cite{Barnich:2009se,Barnich:2011ct,2011JHEP...12..105B} suggested
instead that a larger algebra might be relevant.
In particular,
they suggested adding to the algebra more general
solutions of the conformal Killing equation (\ref{CKV}), in addition to
the six smooth solutions discussed above.  In complex stereographic
coordinates $(z,{\bar z})$ with $z = \cot(\theta/2) e^{i \varphi}$,
the conformal Killing equation reduces to
\be
\partial_z Y^{\bar z} =0, \ \ \ \ \partial_{\bar z} Y^z =0,
\ee
and one can consider solutions $Y^z = Y^z(z)$, $Y^{\bar z} = Y^{\bar
  z}({\bar z})$, where $Y^z$ and $Y^{\bar z}$ are meromorphic
functions of their arguments.  A basis\footnote{Although only real
vector fields $Y^A(\theta^B)$ are physical, for convenience, we use a
complex basis in what follows.  A real basis can be obtained by
taking linear combinations.} of this set of vector fields is
\begin{subequations}
\label{ldef}
\begin{eqnarray}
\label{ldef0}
l_m &=& - z^{m+1} \partial_z,\\
{\bar l}_m &=& -{\bar z}^{m+1} \partial_{\bar z},
\end{eqnarray}
\end{subequations}
for $m \in {\bf Z}$.  Of this infinite basis, the six vector fields
$l_{-1}, l_0, l_1, {\bar l}_{-1}, {\bar l}_0, {\bar l}_1$ are those
discussed after Eq.\ (\ref{CKV}) above that occur in the usual BMS algebra.
The remaining new vector fields are singular and cannot be used to define
smooth, finite diffeomorphisms of the 2-sphere to itself.

The new vector fields have been called ``superrotations'' in the
literature
\cite{Barnich:2009se,Barnich:2011ct,2011JHEP...12..105B,Kapec:2014opa},
since they are generalizations of the six generators of the Lorentz
group.
They might also be called ``superboosts,''
since they are conformal Killing vectors but not Killing vectors on
the 2-sphere, like normal boosts but unlike normal rotations.

What is the physical relevance or utility of the extended\footnote{\label{foot:extendedST}One might think it necessary to include in the algebra
all the vector fields generated by taking
Lie brackets of BMS generators and/or superrotations.
This would yield the algebra summarized in Eqs.\ (4.17) of Ref.\
\cite{Barnich:2010eb}, which contains ``extended supertranslation''
generators of the form (\ref{bmsinfinitesimal}) but where the
function $\alpha$ can contain singularities of the form $z^p {\bar
  z}^q$ with $p$, $q$ negative integers.  These charges associated
with these generators are ill defined \cite{2011JHEP...12..105B}.
However, one could imagine generalizing the definition of asymptotic
flatness by defining a class of solutions which are locally
asymptotically flat on ${\mathscr I}^+$ except for a finite number of
points on the two-sphere which are meromorphic singularities of the kind generated by acting with a finite superrotation \cite{2016arXiv161000639S}.  A diffeomorphism that maps one
such solution onto another has weaker singularities with finite
charges, since one is in effect forbidden from performing two
successive superrotations with the same singular point on the
two-sphere which would change the nature of the singularity.
Thus, the divergent charge integrals computed in  \cite{2011JHEP...12..105B}
might not be a sign of a fatal inconsistency.}  BMS algebra?
There has been some debate in the literature on this issue.
One approach to answering this question is to check whether there are constraints on the
quantum gravity S-matrix associated with the additional symmetries.
Kapec et al.\ \cite{Kapec:2014opa} showed that this is indeed the case
at tree level.
Strominger and Zhibodev \cite{2016arXiv161000639S} showed that
finite superrotations map asymptotically flat spacetimes into a larger
class of spacetimes which are asymptotically flat except at isolated
points which have cosmic string defects.
Finally, another approach is to determine whether there are
well-defined classical conserved quantities for the new symmetries, just as there
are for the standard BMS symmetries.  Barnich and Troessaert followed
this approach in Ref.\ \cite{2011JHEP...12..105B}, where they computed
charge integrals associated with the new generators (\ref{ldef}) with
$|m|>1$ in the Kerr spacetime.
They found that these charges vanish.  In the next section,
we will extend their analysis to more general situations to show that the
charges are finite and to clarify their physical interpretation.

\section{\label{sec:charges} BMS conserved charges}

\subsection{Charges and conservation laws}

We first review the charges and conservation laws associated with the standard BMS group.
There are two types of BMS conservation laws: (i) laws that relate quantities at one cut or crosssection of future null infinity ${\mathscr I}^+$ to another \cite{1981RSPSA.376..585A,1984CQGra...1...15D,Wald:1999wa},
and (ii) laws that relate quantities at past null infinity ${\mathscr I}^-$ to quantities at future null infinity ${\mathscr I}^+$ \cite{PhysRevLett.43.181,Strominger:2013jfa,Pasterski:2015tva}.

Consider the first type of conservation law.
Normally the charges associated with conservation laws can be derived from Noether's theorem.
However, this does not apply to charges associated with BMS generators
at future null infinity, since the associated charges are not actually
conserved because of fluxes of gravitational radiation.
Wald and Zoupas have derived
a generalization of Noether's theorem that allows one to define conserved charges and fluxes
in very general situations of this kind \cite{Wald:1999wa}.  One
obtains for each generator ${\vec \xi}$ a 2-form $\Xi$ on ${\mathscr I^+}$, which depends linearly on ${\vec \xi}$, and
whose integral over any cut (cross section) ${\cal C}$ gives the charge
\be
Q({\cal C}, {\vec \xi}) = \int_{\cal C} \Xi
\label{qformal}
\ee
associated with that cut.  In addition, the exterior derivative $d \Xi$ of the 2-form can be interpreted as a flux that can be integrated over a region ${\cal R}$ of ${\mathscr I^+}$
between two cuts ${\cal C}_1$ and ${\cal C}_2$
to give the change in the charge between two cuts:
\be
\int_{\cal R} d \Xi = Q({\cal C}_2, {\vec \xi}) -Q({\cal C}_1, {\vec \xi}).
\label{qformal1}
\ee
For general relativity, the flux formula had
previously been obtained by Ashtekar and Streubel \cite{1981RSPSA.376..585A},
and the charge associated with a cut had been obtained using a different method
by Dray and Streubel \cite{1984CQGra...1...15D}.

The second type of conservation law is as follows.  Suppose that for a given generator
${\vec \xi}$ of the BMS group acting on ${\mathscr I}^+$
one can identify an associated generator ${\vec \xi}'$ of the BMS
group acting on ${\mathscr I}^-$, with associated 2-form $\Xi'$.
Then one might anticipate a conservation law of the form
\be
\lim_{{\cal D} \to i^0} \int_{\cal D} \Xi' = \lim_{{\cal C} \to i^0}
\int_{\cal C} \Xi,
\label{qformal1a}
\ee
where the first limit to spacelike infinity $i^0$ is taken from the
past along cuts ${\cal D}$ of ${\mathscr I}^-$, and the second limit
to $i^0$ is taken from the future
along cuts ${\cal C}$ of ${\mathscr I}^+$.
Using relations of the form (\ref{qformal1}) on both ${\mathscr I}^-$ and
${\mathscr I}^+$, the conservation law (\ref{qformal1a}) is equivalent
to
\be
\lim_{{\cal D} \to i^-} \int_{\cal D} \Xi'
+ \int_{{\mathscr I}^-} d\Xi'
= \lim_{{\cal C} \to i^+}
\int_{\cal C} \Xi + \int_{{\mathscr I}^+} d\Xi,
\label{qformal1b}
\ee
assuming that the relevant limits exist at future timelike infinity
$i^+$ and past timelike infinity $i^-$.

For the translation subgroup of the BMS group, a method of identifying the
subgroups at ${\mathscr I}^-$ and ${\mathscr I}^+$ was found by
Ashtekar and Magnon-Ashtekar \cite{PhysRevLett.43.181}, together with
an associated conservation law of the form (\ref{qformal1a}) for
4-momentum.  More recently, for the special class of spacetimes
studied by Christodoulou and Klainerman \cite{1993gnsm.book.....C},
Strominger found a method of identifying the two BMS groups
and derived an associated conservation law\footnote{\label{tr}Ashtekar
\cite{Abhay2016} has pointed out that in the
Christodoulou-Klainerman
spacetimes, this conservation law (\ref{qformal1a}) for supermomentum does not yield any
information beyond the conservation of 4-momentum, since the
additional charges all vanish.  This can be seen from
Eqs.\ (2.26) and (2.29) of \cite{Strominger:2013jfa}
and Eq.\ (\ref{ppp}) below.  If the conservation law extends to more general
spacetimes it would yield nontrivial constraints.} of the form (\ref{qformal1b}) for
general generators, in which the boundary
terms at $i^-$ and $i^+$ vanish  \cite{Strominger:2013jfa}.

\subsection{Charges for standard BMS algebra}
\label{sec:standard}

We now turn to the derivation of an explicit expression for general
BMS charges $Q$ in the retarded Bondi coordinates used here.
For simplicity, we
specialize to regions of ${\mathscr I^+}$ which are nonradiative,
and we assume that the leading and subleading
stress-energy components vanish.
It follows that the metric functions $m$ and $C_{AB}$ are
independent of $u$, as discussed in Sec.\ \ref{sec:stationary} above.
Then, for the generator ${\vec \xi}$
given in terms of $\alpha(\theta^A)$ and $Y^A(\theta^B)$ by Eq.\
(\ref{generator}), and for the cut ${\cal C}$ given by $u = u_0$, the
charge is
\begin{eqnarray}
Q &=& \frac{1}{16 \pi} \int d^2 \Omega \bigg[
4 \alpha m
- 2 u_0 Y^A D_A m
+ 2 Y^A N_A
\nonumber \\
&&
- \frac{1}{8} Y^A D_A (C_{BC} C^{BC})
- \frac{1}{2} Y^A C_{AB} D_C C^{BC} \bigg].\ \ \ \
\label{chargeans}
\end{eqnarray}
Note that this charge is independent of $u_0$, i.e., independent of the
cut ${\cal C}$ \footnote{This can be seen by differentiating both sides with respect to $u_0$,
and using the evolution equation (\ref{dotNA}) for the derivative of
$N_A$.  The integral of $Y^A$ against the third and fourth terms on
the right-hand side of Eq.\ (\ref{dotNA}) vanishes since it is proportional to $\int d^2
\Omega \kappa (D^4 + D^6/2) \Psi$, where
$\kappa$ is defined by Eq.\ (\ref{lorentz}) and
$\Psi$ by Eq. (\ref{split}).
This expression vanishes since $\kappa$ is purely $l=1$ while $\Psi$
is purely $l
\ge 2$.}.  This is as expected since the flux $d \Xi$ for all BMS
generators vanishes in nonradiative vacuum regions [see
Eq.\ (\ref{check11}) below].

The formula (\ref{chargeans}) can be obtained from the prescription given after Eq.\
(83) of Wald and Zoupas \cite{Wald:1999wa}.
We decompose the generator ${\vec \xi}$ uniquely as the sum ${\vec \xi} = {\vec
  \xi}_1 + {\vec \xi}_2$ of two generators, where
${\vec \xi}_1$ is tangent to ${\cal C}$ and ${\vec \xi}_2$ is a
supertranslation.
Explicit expressions for the two
generators are
\be
{\vec \xi}_1 = \frac{1}{2} (u-u_0) D_A Y^A \partial_u + Y^A \partial_A
\label{xi1}
\ee
and ${\vec \xi}_2 = (\alpha + \frac{1}{2} u_0 D_AY^A) \partial_u$.
Next, we use the linear dependence of the charge on
${\vec \xi}$  to evaluate the charge as $Q = Q[{\vec
  \xi}_1] + Q[{\vec \xi}_2]$.
The contribution to the
charge from ${\vec \xi}_1$ is given by the integral of the Noether
charge 2-form given by Eq.\ (44) of \cite{Wald:1999wa}
(i.e., the Komar formula).  To get a unique result from this
prescription, Wald explains that from the equivalence class of vector
fields on spacetime that corresponds to the desired BMS generator,
one should choose a representative ${\vec \xi}$ that satisfies
$\nabla_a \xi^a=0$, as proved in Ref.\ \cite{1981JMP....22..803G}.
In fact, an examination of the argument in Ref.\
\cite{1981JMP....22..803G} shows that a sufficient condition for
uniqueness is
$\nabla_a \xi^a = O(1/r^2)$, which is satisfied by the representatives
(\ref{generator}) used here.
Using a vector field of the form (\ref{generator}) associated with the
generator (\ref{xi1}) together with the metric (\ref{metric}) and computing
the integral (92) of Ref.\ \cite{Wald:1999wa} over the surface $u=u_0$ and
$r=r_0$ with $r_0 \to \infty$, we find the third, fourth, and fifth terms
in Eq.\ (\ref{chargeans}) above.
The remaining first and second terms are obtained by inserting the
generator ${\vec \xi}_2$ into the integral (98) of Ref.\ \cite{Wald:1999wa}.
The formula (\ref{chargeans}) was derived by a different
method\footnote{The method used by Barnich and Troessaert is based on
a formula for a variation of the charge which is not
integrable.  This non-integrability issue is resolved in Ref.\
\cite{Wald:1999wa}; however, it does not affect the nonradiative case
considered here.
Formulas similar to Eq.\ (\ref{chargeans}) were derived in Ref.\
\cite{2002hftr.book.....C} [their Eq.\ (6.14)] and Ref.\ \cite{Fujisawa:2015asa},
but those authors found the results $\alpha_1=0$, $\alpha_2 = -1/2$
and $\alpha_1=-1/8$, $\alpha_2 = -2$, respectively, where $\alpha_1$
and $\alpha_2$ are the coefficients of the last two terms in
Eq.\ (\ref{chargeans}), whose values here are $\alpha_1=-1/8$,
$\alpha_2 = -1/2$.  We were unable to uncover the source of the discrepancies.}
by Barnich and Troessaert in Ref.\ \cite{2011JHEP...12..105B}.

We next introduce some notation to describe the different charges.
The quantity (\ref{chargeans}) is a linear function of $\alpha$ and $Y^A$,
and we choose a basis of this vector space as follows.
We parameterize the function $\alpha$ as
\be
\alpha = t^0 - t^i n_i + \sum_{l=2}^\infty \sum_{m=-l}^l \alpha_{lm} Y_{lm},
\label{alphadef}
\ee
where $n_i = (\sin\theta \cos \varphi,\sin \theta \sin \varphi,\cos
\theta)$ and $t^\mu = (t^0,t^i)$ are real parameters.  Similarly we
write $Y^A$ as a linear combination of conformal Killing vectors as
\be
Y^A = \omega^{0i} e^A_{\ i} + \omega^{ij}  e^A_{\ [i} n_{j]},
\ee
where $e^A_{\ i} = D^A n_i$; in Minkowski spacetime, this corresponds
via Eq.\ (\ref{generator}) to the
limiting form of the Killing vector $\omega^{\alpha\beta} x_{[\beta} \partial_{\alpha]}$.
We now define the quantities $P^\mu$, $J^{\mu\nu}$, and ${\cal P}_{lm}$
by
\be
Q = -P^\mu t_\mu + \frac{1}{2} J^{\mu\nu} \omega_{\mu\nu} +
\frac{1}{4\pi}\sum_{l=2}^\infty \sum_{m=-l}^l {\cal P}_{lm}^* \alpha_{lm}.
\label{chargeform}
\ee
Here the four-momentum $P^\mu$ and angular momentum $J^{\mu\nu}$
transform in the normal way under the Lorentz transformation subgroup
of the BMS group given by taking $\alpha =0$ in Eq.\ (\ref{bmsfinite}).
The quantities ${\cal P}_{lm}$ are usually called ``supermomentum''
\cite{McCarthy489,1981RSPSA.376..585A,2015JHEP...03..033B}, although they have also been considered to
be generalizations of angular momentum \cite{2007GReGr..39.2125H}.
We will use the terminology supermomentum since they are conjugate to supertranslations,
they have the same physical
dimension as momentum, and they are invariant under translations
and supertranslations (see Appendix \ref{app:finite}), like normal momentum.

By comparing Eqs.\ (\ref{chargeans}) and (\ref{chargeform}) we see
that the Bondi mass $P^0$ is given by the $l=0$ component of the Bondi
mass aspect $m(\theta^A)$, the linear momentum $P^i$ is given by the
$l=1$ component, and the supermomentum by the higher $l$ components.
Similarly, when $C_{AB} =0$, the angular momentum is given by the $l=1$
component of the angular-momentum aspect $N_A(\theta^B)$, with the
intrinsic angular momentum being encoded in the magnetic parity piece,
and the center-of-mass information being encoded in the electric parity
piece, as discussed in Sec.\ \ref{sec:stationary} above.

The definitions (\ref{chargeform}) are of course dependent on the choice of
Bondi frame: the four-momentum $P^\mu$, angular momentum
$J^{\mu\nu}$, and supermomentum ${\cal P}_{lm}$ transform into one
another under BMS transformations, just as energy and 3-momentum
transform into one another under Lorentz transformations.  See Appendix
\ref{app:finite} for details.

As a check of the charge formula (\ref{chargeans}), in Appendix \ref{app:check}
we compute the flux $d \Xi$ for each generator ${\vec \xi}$,
and verify the expected relation (\ref{qformal1}) between the charges
on two cuts ${\cal C}_1$ and ${\cal C}_2$ and the integral of the flux
over the intervening region ${\cal R}$ of ${\mathscr I^+}$ when the cuts
are in nonradiative vacuum regions.

\subsection{Charges for extended BMS algebra}
\label{sec:charges1}

We now consider the additional symmetries of the extended BMS algebra
discussed in Sec.\ \ref{sec:bmsextended} above.
Should one expect the existence of conserved quantities for singular
symmetry generators ${\vec \xi}$ like superrotations?  The
generalization of Noether's theorem derived in Ref.\ \cite{Wald:1999wa}
remains formally valid, but it is possible that some of the steps in
the argument are invalidated when the vector field ${\vec \xi}$ is not
smooth.  Ideally, one would like to generalize the derivation given
there to the present context.  A simpler alternative, as a first step, is to simply
evaluate the final expression (\ref{chargeans}) for the conserved
charges for the superrotations and see if one obtains a finite result.
This is the approach we will follow here, following Barnich and
Troessaert \cite{2011JHEP...12..105B}.  Clearly a more fundamental
investigation of the applicability of the generalized Noether's
theorem to singular symmetry vector fields is warranted.

As before, we specialize to nonradiative vacuum regions of
${\mathscr I^+}$.
The superrotation charges are obtained from the
general BMS charge
integral (\ref{chargeans}), with $\alpha=0$ and with $Y^A$ taken to be
the superrotation generator $l_m$ given by Eq.\ (\ref{ldef0}) with
$|m| > 1$.  This charge can be written as
\be
Q = \frac{1}{8 \pi} \int d^2 \Omega \, Y^A {\hat N}_A,
\label{sr}
\ee
where
\begin{eqnarray}
{\hat N}_A &=& N_A - u D_A m - \frac{1}{16} D_A (C_{BC} C^{BC})
\nonumber \\
&&- \frac{1}{4} C_{AB} D_C C^{BC}.
\label{hatNdef}
\end{eqnarray}

We next explain why the integral (\ref{sr}) is finite, despite the fact that $Y^A$ is singular.
The integrand in Eq.\ (\ref{chargeans}) is a product of a meromorphic
function of $z$ times a smooth function, and it is a well-known
property of meromorphic functions that such integrals are locally finite.
This result can be understood by expanding the smooth function as a
sum of terms of the form $z^p {\bar z}^q$ with $p,q$ non-negative.
Integrating against a singularity $z^{-m}$ with $m>0$ yields
\[
\int dz d{\bar z} z^{-m} \, z^p {\bar z}^q = \int d\theta \int d\rho \rho^{1 -m + p + q} e^{i \theta(-m + p - q)} \, ,
\]
where $z = \rho e^{i\theta}$.  The angular integral vanishes unless $m  =
p-q$, and then the remaining factor in the integrand is proportional
to $\rho^{1+2 q}$, which is nonsingular\footnote{A similar computation shows that for integrals on the two sphere involving $Y^A$ we can freely integrate by parts despite the singularity, $\int d^2 \Omega D_A (\varphi Y^A)=0$ for any smooth function $\varphi$.}.

Next, we note that we can decompose
${\hat N}_A$ uniquely into electric parity and magnetic parity pieces,
as in Eq.\ (\ref{NAhomsol1}) above.  This gives rise to a
decomposition of the charge (\ref{sr})
into two pieces, which we write as
\be
Q = Q_e + Q_b.
\ee
For the superrotations (\ref{ldef}), we will call the charges $Q_e$
{\it super center-of-mass} charges.
The motivation for this terminology is as follows.
As discussed in Sec.\ \ref{sec:bmsextended}
above, superrotations are generalizations
of Lorentz transformations.  For Lorentz transformations, boosts have
electric type parity, while rotations have magnetic type parity [{\it cf.}\
Eq.\ (\ref{lorentz})].
So it is natural to consider the electric parity pieces of
superrotations to be generalizations of boost symmetries.  Finally,
boost symmetries are conjugate to the center-of-mass piece of angular
momentum.

Similarly, we will call the charges $Q_b$ {\it superspin} charges,
since the magnetic parity pieces
of superrotations can be thought of as generalizations of rotation
symmetries, which are conjugate to intrinsic angular momentum.
For $m = 0, \pm 1$, the charges $Q_e$ and $Q_b$ reduce to the
normal center-of-mass and spin charges discussed in Sec.\
\ref{sec:standard} above.

\subsection{Consistency of charges of extended algebra with fluxes}
\label{sec:consistency}

In Appendix \ref{app:check} we compute the fluxes associated with BMS
generators ${\vec \xi}$ and show that they are consistent with the charge
expression (\ref{chargeans}) in the sense that the conservation law (\ref{qformal1}) is
satisfied for cuts ${\cal C}_1$ and ${\cal C}_2$ in nonradiative regions
of ${\mathscr I}^+$.  We now consider the consistency issue for the
generators ${\vec \xi}$ of the extended algebra.  All of the
computations of Appendix \ref{app:check} continue to apply in this
more general context.  We find from Eq.\ (\ref{qformal2}) that the
conservation law is now not satisfied; there is a discrepancy
proportional to
\begin{eqnarray}
\int du  \int d^2 \Omega
\, Y^A \epsilon_{AB} D^B (D^2 + \frac{1}{2} D^4 ) \Psi,
\label{anssx1}
\end{eqnarray}
where $\Psi$ is the magnetic parity piece of $C_{AB}$, given by Eq.\
(\ref{split}).
This vanishes for BMS generators, for which $Y^A$ is constructed from
$l=1$ harmonics, but not for general superrotations.

What is the explanation for this discrepancy?
The general consistency between the flux and charge formulae derived by Wald
and Zoupas \cite{Wald:1999wa} assumes that the vector fields ${\vec
  \xi}$ are asymptotic symmetry vector fields, that is, vector fields which preserve asymptotic flatness.
This condition is violated
by the superrotation generators used here.
Consistency is restored if we add to the standard BMS flux formula
(\ref{check11}) the quantity\footnote{Alternatively one could restore consistency by using the standard flux (\ref{check11}) and modifying the formula (\ref{chargeans}) for the charge integral by a quantity proportional to the $u$ integral of the magnetic piece of $C_{AB}$.  However this modification would be nonlocal in time.}
\begin{eqnarray}
\frac{1}{32 \pi} \int du  \int d^2 \Omega
\, C_{CE} \, \epsilon_{AB} \epsilon^{CD} D^E D_D D^B Y^A, \ \ \
\label{fix}
\end{eqnarray}
from Eqs.\ (\ref{qformal2}) and (\ref{qformal3}).
The integrand here vanishes identically for standard BMS generators,
but not for superrotations.  We conjecture that the correction (\ref{fix}) gives
the correct flux formula for the extended BMS algebra.
It would be useful to derive this modified flux formula
from a version of the Wald-Zoupas formalism, generalized to accommodate
vector fields of the form used here.
A key element of such a generalization would be an enlarged space
of solutions which are not asymptotically flat but are nevertheless physically relevant,
as described by Strominger and Zhiboedov \cite{2016arXiv161000639S}.

\subsection{Super center-of-mass and superspin charges in stationary regions.}

We now turn to deriving an explicit expression for the charge $Q[l_m]$
associated with the superrotation $l_m$.
We specialize for simplicity
to stationary vacuum regions of ${\mathscr I}^+$, and
to Bondi frames satisfying the constraint (\ref{specialize})
that the angular momentum aspect be non-evolving
(the latter assumption will be relaxed in Appendix
\ref{app:generalize}).
To simplify the computation, we also restrict attention to Bondi
frames\footnote{It might seem that the quantity we are computing is a
  pure gauge effect, since it vanishes in the canonical Bondi gauge.
  This is true locally in time, but as discussed in Sec.\ \ref{sec:memory}
  below,   the change in the charge between two successive stationary regions
  encodes physical, non-gauge information.}
which are close to the canonical Bondi frame (\ref{stat}), so that we can linearize in the
deviation.
In particular, we can neglect the terms quadratic in $C_{AB}$ in Eq.\
(\ref{chargeans}), yielding from Eq.\ (\ref{mconst}) that
\be
Q = \frac{1}{8 \pi} \int d^2 \Omega Y^A N_A.
\label{chargeans1}
\ee
Also, Eq.\ (\ref{NAeqn}) reduces to
\be
D^2 N_A  + N_A= - 3 m_0 D^B C_{AB},
\label{NAeqn2}
\ee
where we have used $m(\theta^A) = m_0$, a constant, from Eq.\ (\ref{mconst}).
We now use the decomposition (\ref{split}) of $C_{AB}$ into electric and
magnetic parity pieces, and use the fact that the magnetic parity
piece vanishes in stationary regions, as shown in Appendix \ref{app:magzero}.
This allows us to solve Eq.\ (\ref{NAeqn2}) to obtain
\be
N_A = - 3 m_0 D_A \Phi/2 + N_A^{l=1},
\label{soln}
\ee
where $N_A^{l=1}$ is a $l=1$, homogeneous solution of Eq.\
(\ref{NAeqn2}) of the type given by Eqs.\ (\ref{NAhomsol1}) and (\ref{NAhomsol2}).
Next, we expand $\Phi$ as
\be
\Phi = \sum_{l \ge 2} \Phi_{lm} Y_{lm},
\label{Phiexpand}
\ee
and
combine Eqs.\ (\ref{ldef0}), (\ref{chargeans1}) and (\ref{soln}) to yield
\be
Q[l_m] = m_0 \sum_{l \ge |m|} \kappa_{lm} \Phi_{l,-m}, \ \ \ \ {\rm
  if}\ \ |m| > 1,
\label{anss}
\ee
where the constants $\kappa_{lm}$ are given by
\be
\kappa_{lm} = - \frac{3}{8} \int_0^\pi d\theta \sin\theta (m -
\cos\theta) \cot^m(\theta/2) Y_{l,-m}(\theta,0).
\label{kappadef}
\ee
This integral evaluates to ${\hat \kappa}_{lm}$ for $m \ge 2$, and to $(-1)^{l+1} {\hat \kappa}_{l(-m)}$ for $m \le -2$, where
\be
{\hat \kappa}_{lm} = - \frac{3}{8} \sqrt{ \frac{ (2 l + 1) (l+m)!}{\pi (l-m)!}}
\frac{1}{(l^2 + l -2) \Gamma(m-1)}.
\ee
These constants are finite even for $m < 0$.
The charge $Q[{\bar l}_m]$ is given by a similar expression but with
$\Phi_{l,-m}$ replaced by $\Phi_{lm}$ in Eq.\ (\ref{anss})
and with $Y_{l,-m}$ replaced by $Y_{lm}$ in Eq.\ (\ref{kappadef}).

We note that the final result (\ref{anss}) comes purely from the
electric parity piece of the expression (\ref{soln}), since the second
term in that expression does not contribute.
Hence the charge (\ref{anss}) is a super center-of-mass charge, $Q=
Q_e$, while the superspin charge $Q_b$ is vanishing.

The final result is that
the super center-of-mass charges $Q_e[l_m]$ and $Q_e[{\bar l}_m]$ give information about the tensor $C_{AB}$, via Eqs.\ (\ref{split}), (\ref{Phiexpand}),
and (\ref{anss}).
We note that the information is incomplete, since one cannot reconstruct
$C_{AB}$ from these charges.


The computations in this subsection of the charges (\ref{anss}) were specialized to Bondi frames
obeying the constraint (\ref{specialize}) that the angular momentum aspect be non-evolving.
While such frames always exist in stationary regions, they are not the most general frames.
In appendix \ref{app:generalize} we generalize the computations to
remove this constraint, while retaining the assumption of
linearization about the canonical Bondi frame.
The result is that there are modifications to the charges for $|m|\le
2 $ but not for higher values of $|m|$.  In particular the superspin
charges are no longer vanishing, for $|m| \le 2$.

For more general Bondi frames in stationary
regions, the superspin charges $Q_b$ need not vanish.
However, these charges do not contain any information
not already contained in the standard BMS charges and the super
center-of-mass charges $Q_e$.
This follows from the fact
that all the superrotation charges vanish in the canonical Bondi frame
in stationary regions, from Eqs.\ (\ref{stat}) and (\ref{chargeans}),
and from considering the number of free
functions in the general BMS transformation
(\ref{bmsfinite}) to an arbitrary Bondi frame.

\subsection{Changes in super center-of-mass and superspin charges in nonradiative to nonradiative transitions}

We now turn to considering the super center-of-mass and superspin
charges in more general, nonstationary but still nonradiative
situations.  In these situations, both types of charge are finite,
by the argument given in Sec.\ \ref{sec:charges1} above.
They are also independent, and it is not possible to compute them in
terms of the strain tensor $C_{AB}$.

For the superspin charges $Q_b$, we now derive a formula
for the change in the charges in a transition from an early
nonradiative region at $u=u_1$ to a later nonradiative region at $u=u_2$.
Taking the magnetic part of Eq.\ (\ref{hatNAdot})
and integrating with respect to $u$ [or equivalently using the
corrected flux given by Eqs.\ (\ref{fix}) and (\ref{check11})]
and combining with Eq.\ (\ref{sr})
gives the total change
\begin{eqnarray}
Q_b(u_2)&& - Q_b(u_1)= -\int d^2 \Omega \int du Y^A ( {\hat T}_{uA}^m +
{\cal T}_{uA}^m ) \nonumber \\
&& + \frac{1}{64 \pi} \int d^2 \Omega \int du \, Y^A \epsilon_{AB} D^B D^2
(D^2 + 2) \Psi.\ \ \ \ \ \ \ \
\label{ansd}
\end{eqnarray}
Here the superscript $m$ means ``magnetic part of'', and the quantity
${\cal T}_{uA}$ is given by
\begin{eqnarray}
{\cal T}_{uA} &=& \frac{1}{64 \pi} \bigg[
3 N_{AB} D_C C^{BC} -3 C_{AB} D_C N^{BC}
\nonumber \\
&&
-  D_B C_{AC} N^{BC}
 +  D_B N_{AC} C^{BC} \bigg],
\label{calTdef}
\end{eqnarray}
a kind of gravitational wave angular momentum flux\footnote{Note
  that this flux differs from the gravitational wave angular momentum
  flux defined in Eq. (2.3) of Ref.\ \cite{Pasterski:2015tva}.  That
  flux
characterizes the evolution of the angular momentum aspect, but not
the radiated angular momentum.  The two fluxes differ by a total derivative with respect to $u$.}.
This result is consistent with Pasterski, Strominger and Zhiboedov
\cite{Pasterski:2015tva}, who argued that the conserved
quantities\footnote{They derived a conservation law
of the form (\ref{qformal1b})
for gravitational
  scattering from past null infinity to future null infinity for
Christodoulou Klainerman spacetimes.}
associated with the superrotation symmetries are of the form\footnote{The difference between the angular momentum flux definition (\ref{calTdef}) and that used in \cite{Pasterski:2015tva}
implies that the charges here and there do not coincide; however the two conservation laws
are equivalent in the sense that each can be derived from the other.}
(\ref{ansd}).
Here we extend their arguments to also include the electric pieces
(\ref{anss}) of
the superrotation charges.

\section{\label{sec:interpretation} Physical interpretation of BMS and extended BMS charges}

In this final section, we make some remarks about the physical significance of
BMS charges and, in particular, about how they can be measured.

\subsection{General considerations}

For all symmetry generators ${\vec \xi}$ and cuts ${\cal C}$ of future
null infinity ${\mathscr I^+}$, the charges $Q[{\vec \xi},{\cal C}]$ are defined in
terms of integrals over ${\cal C}$, which can be evaluated in terms a
limit of integrals over finite 2-surfaces that tend to ${\cal C}$.
It follows that the charges can in principle be measured by
a collection of observers distributed over a 2-surface near future
null infinity, who each make local measurements of the spacetime
geometry in their vicinity and of their motion and orientation
relative to their neighbors, who then communicate this information
to one another, and who finally process it in a suitable way.
Thus, in principle, the charges are measurable quantities.
Of course, it would be useful to understand in more detail
how to specify a local\footnote{By this we mean to exclude, for
  example, demanding that observers be stationary with respect to
  retarded Bondi coordinates, which would be a nonlocal requirement.}
operational prescription\footnote{Such a prescription can be given
for the Poincar\'e charges in stationary situations \cite{Flanagan:2014kfa,Flanagan:2016bb}.}
for such measurements.  The formalism of rigid quasilocal frames of
Refs.\
\cite{2009CQGra..26c5015E,2013arXiv1307.1914E,2013CQGra..30s5019E,2014arXiv1402.1443M}
might be useful for this purpose, as well as the
covariant-conformal-completion formalism for asymptotically flat
spacetimes \cite{1984ucp..book.....W}.

Specifying an asymptotic Bondi frame allows observers to establish a
convention for labeling the various charges.
By an asymptotic Bondi frame, we mean a choice of coordinates
$(u,\theta^A)$ on ${\mathscr I^+}$ for which the spacetime metric takes
the form (\ref{metric})---which is unique up to BMS transformations.
A Bondi frame is determined up to an overall $SO(3)$ rotation
by specifying a single cross section or
cut of ${\mathscr I^+}$.
One can equivalently think of a Bondi frame as an
equivalence class of coordinate systems on spacetime whose asymptotic
limits coincide in a suitable way, or as a class of asymptotic
observers who adjust their relative motions, clocks and orientations in such
a way as to establish an approximate consistent convention for specifying the
results of asymptotic measurements.\footnote{There is a close analogy to local
  Lorentz frames, which can be thought of as a specification of a
  set of orthonormal basis vectors at a point in spacetime, an
  equivalence class of local coordinate systems, or a class of observers who
  adjust their motions, clocks and orientations in order to establish an approximate convention for
  specifying the results of measurements of components of tensors near
  that point.}

\subsection{Nonradiative vacuum regions of future null infinity}

In nonradiative vacuum regions of ${\mathscr I}^+$, the Bondi mass aspect
$m = m(\theta^A)$ and shear $C_{AB}(\theta^C)$ are independent of $u$,
as argued in Sec.\ \ref{sec:stationary} above.
Moreover the angular momentum aspect has the form
$N_A(u,\theta^B) = {}^0N_A(\theta^B) +  {}^1N_A(\theta^B) u$, from Eq.\ (\ref{dotNA}).
These functions
can in principle be extracted from measurements of
the asymptotic components of the Weyl tensor, listed in Appendix
\ref{app:weyl}; see Ref.\ \cite{raphael} for details.

The Bondi mass aspect $m(\theta^A)$ encodes the Bondi 4-momentum and supermomentum, as in Eq.\ (\ref{ppp}) below.
In stationary regions, the supermomentum does not contain any additional information aside from the Bondi 4-momentum, from Eqs.\
(\ref{mstationary}) and (\ref{mboost}).  However in more general nonradiative regions it does.  For example, for the linearized gravitational field of two point particles which have a relative boost, one can extract from the supermomentum the individual 4-momenta of the two particles.
The conservation laws associated with supermomentum can be described as
a separate conservation law for energy at every angle, as explained by
Strominger \cite{Strominger:2013jfa}.

Similarly the combination ${\hat N}_A$
of the angular momentum aspect and strain tensor given by Eq.\ (\ref{hatNdef}) encodes the super center-of-mass and superspin charges via
Eq.\ (\ref{sr}).  The super center-of-mass charges encode
the supertranslation that relates the given BMS frame to the BMS center-of-mass frame; they can be set to zero using a supertranslation.
In stationary regions the superspin charges encode just the spin of the spacetime.  However in more general, nonradiative regions they contain more information, like the supermomentum.
For example, for the linearized gravitational field of two spinning point particles which have a relative boost, one can extract from the superspin charges the individual spins of the two particles.
The conservation laws associated with superspin could be described
as a separate conservation law for angular momentum at every angle [{\it cf.}\ Eq.\ (\ref{h4}) below].

\subsection{\label{sec:memory} Relation to gravitational-wave memory}

Gravitational-wave memory is the relative displacement of initially
comoving observers caused by the passage of a burst of gravitational
waves \cite{1974SvA....18...17Z,1987Natur.327..123B,braginskii1985kinematic}.
There is a well-known close relation between gravitational-wave
memory for observers near future null infinity and the BMS group:
the supertranslation that relates the canonical Bondi frame of an
initially nonradiative region to that of a final nonradiative region
encodes the observed memory effect
\cite{1966JMP.....7..863N,1981JMP....22..803G,2007GReGr..39.2125H}.
See Strominger et al.\ \cite{Strominger:2014pwa} for a recent clear
exposition of this relation in the retarded Bondi coordinates used
here.  The memory/supertranslation effect can also be characterized in
a gauge-invariant but nonlocal way in terms of a generalized holonomy
around a suitable closed loop in spacetime near ${\mathscr I^+}$ \cite{Flanagan:2014kfa}.

Here we point out a new aspect of this story: a close
correspondence between the two different infinite families of extended
BMS charges (supermomentum and super center-of-mass) and the two
different types of memory (ordinary and null \cite{Bieri:2013ada}).

Consider a spacetime in which the flux of energy to future null
infinity vanishes in the vicinity some early retarded time $u_1$, so that the news
tensor $N_{AB}$ and stress-energy tensor vanish there.  Suppose that there
is subsequently a burst of gravitational waves and/or matter energy
flux to infinity, and that the fluxes vanish again in the vicinity of some later
retarded time $u_2$.  Freely falling, initially comoving adjacent observers
near infinity can measure their net relative displacement, and as
shown in Ref.\ \cite{Strominger:2014pwa}, to leading order in $1/r$
this displacement is encoded in the change
\be
\Delta C_{AB} = C_{AB}(u_2) - C_{AB}(u_1),
\label{DeltaC}
\ee
of the tensor $C_{AB}$.  Thus, we will identify the change (\ref{DeltaC}) as
the gravitational-wave-memory observable.\footnote{From Eq.\ (\ref{deltaCAB})
the change $\Delta C_{AB}$ is invariant under supertranslations and
transforms just under the Lorentz group.}

The observable change (\ref{DeltaC}) can be decomposed into electric parity
and magnetic parity pieces, as in Eq.\ (\ref{split}):
\be
\Delta C_{AB} = (D_A D_B - \frac{1}{2} h_{AB} D^2) \Delta \Phi
+ \epsilon_{C(A} D_{B)} D^C \Delta \Psi,
\label{split1}
\ee
where $\Delta \Phi = \Phi(u_2) - \Phi(u_1)$ and $\Delta \Psi = \Psi(u_2) - \Psi(u_1)$.
These two pieces can in principle be measured by surrounding a source
of gravitational waves with a collection of observers distributed on a
2-sphere, having them each measure the gravitational-wave memory, and
then decomposing the resulting function on the 2-sphere into electric
and magnetic pieces, as discussed by
Winicour \cite{2014CQGra..31t5003W,2016CQGra..33q5006M}.  This would be analogous to
measurements of E and
B modes of the cosmic-microwave-background polarization.
We now discuss these two pieces separately.

\subsubsection{Electric parity piece of shear}
\label{sec:ee}

We can compute the electric parity piece $\Delta \Phi$ as follows.
Following Ref.\ \cite{Strominger:2014pwa}, we substitute the decomposition
(\ref{split}) into the evolution equation (\ref{dotm}) for the Bondi mass
aspect and integrate from $u_1$ to $u_2$.
The result is
\be
\Delta m = -4 \pi \Delta {\cal E} + {\cal D} \Delta \Phi.
\label{ll}
\ee
Here $\Delta m = m(u_2) - m(u_1)$ is the change in the Bondi mass
aspect, ${\cal D}$ is the angular
differential operator
\be
{\cal D} = D^2/4 + D^4/8,
\label{calDdef}
\ee
and
\be
\Delta {\cal E} =  \int_{u_1}^{u_2} du \left[  {\hat T}_{uu} +
  \frac{1}{32 \pi} N_{AB} N^{AB} \right]
\label{DeltaE}
\ee
is the total energy radiated per unit solid angle in
either matter or gravitational waves.  Next, we act on both sides
of Eq.\ (\ref{ll}) with
the projection operator ${\cal P}$ that sets to zero the $l=0,1$
pieces of functions on the sphere, and by the inverse\footnote{See
  footnote \ref{foot:ProjL01} above.} of the operator
${\cal D}$.  Using ${\cal P} {\cal D} = {\cal D}$ the result is
\cite{Strominger:2014pwa}
\be
\Delta \Phi = {\cal D}^{-1} {\cal P} \Delta m + 4 \pi {\cal D}^{-1} {\cal P} \Delta
  {\cal E} .
\label{bieri}
\ee

The left-hand side of this equation is the observable, the (electric parity
piece of) the gravitational-wave memory.  The second term on the
right-hand side is what Bieri and Garfinkle called the {\it null
  memory}, the piece of the memory that is computable directly in
terms of fluxes of energy\footnote{Bieri and Garfinkle worked in the
  context of linearized gravity, so they did not have the gravitational-wave
energy-flux term in Eq.\ (\ref{DeltaE}).  This term was originally computed in
vacuum by Christodoulou \cite{PhysRevLett.67.1486} who called the effect
``nonlinear memory.''  The formula with both matter and gravitational-wave
fluxes was derived in \cite{Strominger:2014pwa}.} to future null infinity.
The first term is what Bieri and Garfinkle called {\it ordinary
  memory}, the kind originally discussed by Zel'dovich
\cite{1974SvA....18...17Z}, which is computable from the change in the
asymptotic component (\ref{weyl1}) of the Weyl curvature tensor between early and late times.

We see from Eq.\ (\ref{bieri}) that
the ordinary memory is
reflected in the $l\ge 2$ components of the Bondi mass aspect
$m(\theta^A)$, or equivalently the supermomenta ${\cal P}_{lm}$, from
Eq.\ \eqref{ppp}.  The total, ordinary plus null memory is encoded
in the shear tensor, or $\Delta \Phi$. In the special case of
stationary-to-stationary transitions this is in turn (partially) encoded in the
super-center-of-mass charges, from Eq.\ (\ref{anss}).

\subsubsection{Magnetic parity piece of shear and spin memory}
\label{sec:bb}

Turn now to the magnetic parity piece $\Delta \Psi$ of the shear.
For stationary-to-stationary transitions, it follows from the result
of Appendix \ref{app:magzero} that $\Delta \Psi$ vanishes.
For more general nonradiative-to-nonradiative transitions,
it is known
that $\Delta \Psi$ vanishes in the context of linearized gravity
\cite{Bieri:2013ada,2016CQGra..33q5006M}.
We conjecture that $\Delta \Psi$ also vanishes in full general
relativity for such transitions.
If this conjecture is true, then there is no magnetic piece of normal
gravitational-wave memory.

However, there is another observable that will be generically
nonvanishing, the time integral
over the burst of gravitational waves of the magnetic piece of the
shear, or
\be
\int du \Psi.
\label{spinmemory}
\ee
This constitutes a new type of gravitational wave memory, {\it spin
  memory}, discovered by Pasterski, Strominger and Zhiboedov
\cite{Pasterski:2015tva}.  It can be measured by observers who monitor
the time dependent gravitational wave strain, integrate that quantity
with respect to time, and decompose on a 2-sphere to extract the
magnetic parity part.  The time integral of the shear can
alternatively be measured in principle
by measuring the mapping between initial relative displacement and
velocity,
and final relative displacement and velocity for a pair of adjacent
freely falling test masses \cite{abe1}.
Finally, the spin memory observable (\ref{spinmemory})
can also be measured using Sagnac interferometers by a certain class
of accelerated observers \cite{Pasterski:2015tva}.

Just as for normal (electric parity) memory, spin memory can be
decomposed into null and ordinary pieces.
Following \cite{Pasterski:2015tva} we integrate the magnetic piece of
Eq.\ (\ref{hatNAdot}) with respect to $u$ and contract with
$\epsilon^{AC} D_C$ to obtain
\be
\Delta ( \epsilon^{AC} D_C {\hat N}_A)
= - 8 \pi \epsilon^{AC} D_C \Delta {\cal E}_A + D^2 {\cal D} \int du \Psi.
\label{h4}
\ee
Here ${\cal D}$ is given by Eq.\ (\ref{calDdef}),
\be
\Delta {\cal E}_A =  \int_{u_1}^{u_2} du \left[  {\hat T}_{uA} +
  {\cal T}_{uA} \right]
\label{DeltaEA}
\ee
is the total angular momentum radiated per unit solid angle in
either matter or gravitational waves, and ${\cal T}_{uA}$ is given by
Eq.\ (\ref{calTdef}).  It follows that
\begin{eqnarray}
\int du \Psi &=& {\cal D}^{-1} D^{-2} {\cal P} \Delta ( \epsilon^{AC}
D_C {\hat N}_A) \nonumber \\
&&+ 8 \pi {\cal D}^{-1} D^{-2} {\cal P} \epsilon^{AC} D_C \Delta {\cal E}_A.
\end{eqnarray}
The second term here is null spin memory, computable in terms of the
flux of angular momentum to null infinity.  The first term is ordinary
spin memory, computable from the changes in the asymptotic components
of the Weyl tensor, or, equivalently, from changes in the superspin
charges.

\subsection{\label{sec:evolution} BMS charges as black-hole hair}

The BMS charges we have been discussing are universal, applying to
any kind of isolated object in an asymptotically flat spacetime.  In
particular, they apply to black holes.  In this context, they can be
thought of as a kind of black-hole hair, as pointed out by Strominger
and Zhiboedov (SZ) \cite{Strominger:2014pwa}, who dubbed them ``soft
hair''.
SZ discussed how this hair could be measured
in terms of a gravitational-wave-memory observation.  Here we expand on that
description to clarify how the no-hair theorems \cite{1996CLNP....6.....H} are evaded.

The supermomentum charges ${\cal P}_{lm}$ characterize departures from
stationarity [this follows from Eqs.\ (\ref{mstationary}), (\ref{chargeans}),
and (\ref{chargeform}) above], so they vanish\footnote{More precisely, in a
  stationary region of ${\mathscr I^+}$ the supermomentum components ${\cal P}_{lm}$  need not vanish, depending on the choice of Bondi frame.   However, they are
  determined by the Bondi 4-momentum, so they contain no additional information.} for black holes once they settle down to a stationary state.  Thus, black
holes do not admit supermomentum hair.  The same is true for superspin
charges.

The super center-of-mass charges (i.e. the electric parity piece of
the shear tensor $C_{AB}$) are closely
analogous to the angles
$(\theta,\varphi)$ that specify the direction of the black hole's
intrinsic angular momentum ${\bf S}$.  Do those angles constitute
black hole hair?  Clearly, they do not give information about intrinsic
properties of the black hole, since they merely reflect an orientation
with respect to an arbitrarily chosen asymptotic reference frame.
On the other hand, if one considers observations at more than one
time, and if the black hole accretes some angular momentum, then
angle $\Delta \Theta$ by which the orientation changes
between early and late times is a physical property of
the black hole, independent of any choice of reference frame.
In addition, even if one restricts attention to one instant of time,
in quantum mechanics one can have superpositions of different angular-momentum
eigenstates, and the existence of a nontrivial superposition
is again a physical property of the black hole, independent of any
choice of reference frame.

The super center-of-mass hair is exactly analogous.
In the classical theory, at one instant of time, they do not give any
information about intrinsic properties of the black hole.  Instead, they
give information about properties of the black hole relative to an arbitrarily
chosen asymptotic Bondi frame, and those properties can be made to
vanish with a suitable choice of Bondi frame in stationary situations
[cf.\ Eq.\ (\ref{CABstationary}) above].  Hence the no-hair theorems are not
violated.  On the other hand, if one considers measurements made at
two different times at which the black hole is stationary,
then the changes (\ref{DeltaC}) in the charges give nontrivial
physical information, independent of any choice of reference frame.
This information is the gravitational-wave memory/supertranslation/generalized
holonomy, as explained by SZ.  Finally, if one restricts attention to
one instant of time, one can have nontrivial superpositions of super
center-of-mass eigenstates, and the existence of a nontrivial
superposition is a physical property of the black hole, independent of
any choice of reference frame.  To produce such superpositions one can
throw into a black hole matter that is in a superposition of two
states, one state for which the gravitational-wave emission associated with
the accretion produces a net gravitational-wave memory, and one state
for which the net memory is zero.
Thus in quantum gravity, the set of quantum states
associated with low energy, asymptotic degrees of freedom of the black hole
is richer than what would be expected from the classical theory locally in time.

\section{Conclusions}

In this paper, we have investigated the suggestion of Refs.\
\cite{Banks:2003vp,Barnich:2009se,Barnich:2011ct,2011JHEP...12..105B}
that the BMS symmetry algebra be extended.
While we have found that some of the symmetry generators of the
extended algebra have conserved charges that are finite
and are associated with gravitational-wave memory, there are several
outstanding puzzles and open issues:

\begin{itemize}


\item Consistency of the superspin charges with fluxes requires a correction to
  the standard formula for the flux associated with the BMS generator.
  It would be useful to derive this correction from first principles.

\item We computed the charges only in a certain regime where some
nonlinearities could be neglected [cf.\ the discussion before Eq.\
(\ref{chargeans1}) above] and only in stationary, vacuum regions of
${\mathscr I^+}$.  It might be interesting to
investigate the properties of the charges more generally.

\item The new charges capture some but not all of the information
  associated with the observable gravitational-wave memory---cf.\ the
  discussion after Eq.\ (\ref{kappadef}) above.
This suggests that yet larger symmetry algebras might be relevant.

\end{itemize}

A summary the status of the various charges and results discussed in this paper
is given in Table \ref{tab:summary}.

\begin{table*}[t]
\centering
\footnotesize
\begin{tabular}{|C{2.5cm}|C{2.5cm}|C{2cm}|C{2cm}|C{2cm}|C{2.5cm}|C{3cm}|}
\hline
Charge&Symmetry&
Metric function in which encoded& Interpretation of conservation law & Status of conservation law
from one cut of ${\mathscr I}^+$ to another & Status of conservation law for gravitational scattering
from ${\mathscr I}^-$ to ${\mathscr I}^+$ &
Relation to gravitational wave memory    \\
\hline
\hline
Supermomentum & Supertranslations (standard BMS algebra) &
Bondi mass aspect $m(\theta^A)$ &
Energy conservation at every angle \cite{Strominger:2013jfa}
 & Established in \cite{1981RSPSA.376..585A,1984CQGra...1...15D,Wald:1999wa}
& Established in \cite{Strominger:2013jfa}
for Christodoulou-Klainerman spacetimes \cite{1993gnsm.book.....C}.
In that context supermomentum conservation contain no information
beyond 4-momentum conservation\footnote{See footnote \ref{tr}.}.
Not yet established more generally.
& Changes in supermomentum encode ordinary \cite{Bieri:2013ada} piece of normal gravitational wave memory [Sec.\ \ref{sec:ee}]
\\
\cline{1-7}
Superspin & Magnetic parity piece of superrotations (extended BMS algebra) &
Magnetic parity piece of angular momentum  aspect $N_A(\theta^B)$ &
Angular momentum conservation at every angle \cite{Pasterski:2015tva}
 & Established in \cite{Pasterski:2015tva}
& Established in \cite{Pasterski:2015tva}
for Christodoulou-Klainerman spacetimes.
Not yet established more generally.
& Changes in superspin encode ordinary \cite{Bieri:2013ada} piece of gravitational wave spin memory \cite{Pasterski:2015tva} [Sec.\ \ref{sec:bb}]
\\
\cline{1-7}
Super center-of-mass (also called {\it soft hair} in the context of black holes
\cite{2016PhRvL.116w1301H})
 & Electric parity piece of superrotations (extended BMS algebra) &
Electric parity piece of angular momentum  aspect $N_A(\theta^B)$ &
Center-of-mass conservation
at every angle [Sec.\ \ref{sec:charges1}]
 & Established here [Sec.\ \ref{sec:consistency}]
& Not yet established.
& Changes in super center-of-mass charges encode the total (ordinary plus null) normal
gravitational wave memory for stationary-to-stationary transitions [Sec.\ \ref{sec:ee}]
\\
\hline\hline
\end{tabular}

\caption{
A summary of the charges discussed in this paper, the corresponding symmetries,
the status of the corresponding conservation laws, and the relations to gravitational wave memory.}
\label{tab:summary}
\end{table*}

\normalsize

\acknowledgments
We thank David Garfinkle, Abhay Ashtekar and Blagoje Oblak
for helpful conversations, and Leo Stein for helpful discussions and for evaluating the
integral (\ref{kappadef}).  We thank Andy Strominger, Sabrina Pasterski and Thomas Maedler
for helpful correspondence, and an anonymous referee for insightful
comments.  This research was supported in part by NSF grants
PHY-1404105 and PHY--1068541.

\appendix
\section{\label{app:weyl} Weyl tensor}

For vacuum solutions, the leading-order components of the Weyl tensor in the retarded Bondi coordinates
are as follows.  We define the basis of vector fields
\be
{\vec e}_{\hat u} = \partial_u, \ \ \ {\vec e}_{\hat r} = \partial_r, \ \ \  {\vec e}_{\hat A}  = \frac{1}{r} \partial_A,
\ee
which is asymptotically orthonormal as $r \to \infty$.
The leading-order, $O(1/r)$ components of the Weyl tensor on this basis are Petrov type IV from the peeling theorem.
They are given by
\be
C_{{\hat u}{\hat A}{\hat u}{\hat B}} = - \frac{1}{2 r} {\ddot C}_{AB}
\ee
with all other components (except those related to these by symmetries) vanishing at this order.
The subleading-order, $O(1/r^2)$ components are given by
\begin{subequations}
\begin{eqnarray}
C_{{\hat u}{\hat r}{\hat u}{\hat A}} &=& - \frac{1}{2 r^2} D^B {\dot C}_{AB}, \\
C_{{\hat u}{\hat A}{\hat u}{\hat B}} &=& \frac{1}{4 r^2} \left[ D^2 {\dot C}_{AB} - 2 {\dot C}_{AB} - h_{AB} C_{CD} {\ddot C}^{CD}\right]\nonumber \\
&& \\
C_{{\hat u}{\hat A}{\hat B}{\hat C}} &=& - \frac{1}{r^2} D_{[B} {\dot C}_{C]A}.
\end{eqnarray}
\end{subequations}
At order $O(1/r^2)$, the Weyl tensor is Petrov type III.
The remaining components scale as $C_{{\hat u}{\hat r}{\hat u}{\hat r}} \sim C_{{\hat u}{\hat r}{\hat A}{\hat B}} \sim C_{{\hat A}{\hat B}{\hat C}{\hat D}} \sim r^{-3}$ and $C_{{\hat u}{\hat r}{\hat r}{\hat A}} \sim C_{{\hat r}{\hat A}{\hat r}{\hat B}} \sim C_{{\hat r}{\hat A}{\hat B}{\hat C}} \sim r^{-4}$.
In particular we have
\be
C_{{\hat u}{\hat r}{\hat u}{\hat r}} = -  \frac{1}{r^3} \left[ 2 m + \frac{1}{4} C_{AB} N^{AB} \right] + O\left( \frac{1}{r^4} \right)
\label{weyl1}
\ee
and
\be
C_{{\hat u}{\hat r}{\hat r}{\hat A}} = \frac{N_A}{r^4} + O\left( \frac{1}{r^5} \right).
\label{weyl2}
\ee

\section{\label{app:finite} Transformation properties of charges under finite BMS transformations}

The transformation properties of the charges
under finite BMS transformations can be derived
from the definitions (\ref{bmsfinite}), (\ref{bmsinfinitesimal}), and (\ref{qformal});
see, for example, Appendices C.5 and C.6 of Chrusciel {\it et.\ al.}
\cite{2002hftr.book.....C}.
Here we restrict attention to vacuum nonradiative regions of ${\mathscr I^+}$ so
that we can neglect the cut dependence of the charges.
Consider a finite BMS transformation $\psi : {\mathscr I^+} \to {\mathscr I^+}$ of the form (\ref{bmsfinite}).  It can be parameterized in
terms of a map $\varphi : S^2 \to S^2$ of the 2-sphere to itself and
a function $\beta$ on the 2-sphere [denoted by $\alpha$ in Eq.\
(\ref{bmsfinite})].
The BMS generator ${\vec \xi} = (\alpha, Y^A)$ given by Eq.\ (\ref{bmsinfinitesimal})
is mapped by the pullback to $\psi_* {\vec \xi} = ({\hat \alpha}, {\hat Y}^A)$,
where ${\hat Y}^A = \varphi_* Y^A$,
\be
{\hat \alpha} =  \omega_\varphi \, \varphi_* \alpha + \frac{1}{2} \beta D_A {\hat Y}^A - {\hat Y}^A D_A \beta,
\ee
and $\omega_\varphi$ is defined by $\varphi_* h_{AB} =
\omega_{\varphi}^{-2} h_{AB}$.  For rotations $\omega_\varphi=1$,
while for a boost with rapidity parameter $\eta$ in the direction
${\bf m}$, $\omega_{\varphi}({\bf n}) = \cosh \eta + \sinh \eta {\bf
  n} \cdot {\bf m}$.  The charges transform according to\footnote{We
  use $\psi^{-1}$ instead of $\psi$ in this transformation law in order
to agree with convention of the linearized analysis of Sec.\ \ref{sec:bmsgroup}
above.  The charges are invariant under ${\vec \xi} \to \psi_* {\vec
  \xi}, g_{ab} \to \psi_* g_{ab}$, so the convention $g_{ab} \to
\psi_* g_{ab}$ used there is equivalent to the convention ${\vec \xi}
\to \psi_*^{-1} {\vec \xi}$ used here.}
\be
Q[{\vec \xi}] \to {\bar Q}[{\vec \xi}]= Q[\psi^{-1}_* {\vec \xi}].
\ee

For boosts and rotations the results are as follows.  The 4-momentum
and angular momentum transform in the standard way:
\be
{\bar P}^\alpha = \Lambda^\alpha_{\ \beta} P^\beta, \ \ \ \ \ {\bar J}^{\alpha\beta} = \Lambda^\alpha_{\ \mu} \Lambda^\beta_{\ \nu} J^{\mu\nu}.
\ee
Here $\Lambda^\alpha_{\ \beta}$ is the Lorentz transformation that is
naturally associated with $\varphi$, which can be obtained by demanding that
the action of the Lorentz transformation on the set of null directions
$(1, {\bf n})$ coincide with that of $\varphi^{-1}$.\footnote{Formally the
  Lorentz transformation $\Lambda^\alpha_{\ \beta}$ as well as
  $P^\alpha$ and $J^{\alpha\beta}$ are tensors over the 4-parameter
  translation subgroup of the BMS group, which has a flat $(-,+,+,+)$
  metric \cite{bobg}.}
For the supermomentum, it is more convenient to give the
transformation law in terms of the Bondi mass aspect, which encodes
the supermomenta according to [cf.\ Eqs.\ (\ref{chargeans}),
(\ref{alphadef}), and (\ref{chargeform})]
\be
m(\theta^A) = P^0 + 3 P_i n^i + \sum_{l \ge 2} \sum_m {\cal P}_{lm}
Y_{lm}.
\label{ppp}
\ee
The transformation law is
\be
{\bar m}(\theta^A) = m[\varphi(\theta^A)] \omega_{\varphi}(\theta^A)^{-3}.
\label{mboost}
\ee
Finally the tensor $C_{AB}$ which encodes the super center-of-mass charges transforms as
\be
{\bar C}_{AB} = \omega_\varphi \, \varphi_* C_{AB}.
\label{dd}
\ee

Consider next translations and supertranslations, which are
parameterized by the function $\beta$.  The 4-momentum and
supermomentum are invariant under these transformations.
The angular momentum transforms as ${\bar J}_{\mu\nu} = J_{\mu\nu} +
\delta J_{\mu\nu}$ with
\begin{subequations}
\begin{eqnarray}
\delta J_{ij} &=&  \frac{1}{2\pi} \int d^2 \Omega\,  m \, e^A_{[i} n_{j]} D_A
\beta, \\
\delta J_{0i} &=& - \frac{1}{4\pi} \int d^2 \Omega\,  \beta ( e^A_{\ i}
D_A m - 3 n_i m). \\
\nonumber
\end{eqnarray}
\end{subequations}
By using Eq.\ (\ref{ppp}) these angular momentum changes can be expressed in
terms of the 4-momentum and supermomentum.  For a normal translation
with $\beta = t^0 - t^i n_i$ we recover the standard transformation
law $\delta J_{\mu\nu} = -t_\mu P_\nu + t_\nu P_\mu$.
Finally the tensor $C_{AB}$ transforms according to the same transformation law as in the linearized case:
\be
{\bar C}_{AB} = C_{AB} - 2 D_A D_A \beta + h_{AB} D^2 \beta.
\label{st11}
\ee

\section{Verification of flux conservation law for standard BMS algebra}
\label{app:check}

In this appendix we compute the flux $d \Xi$ for each generator
${\vec \xi}$ of the standard BMS algebra,
and verify the expected relation (\ref{qformal1}) between the charges
on two cuts ${\cal C}_1$ and ${\cal C}_2$ and the integral of the
flux over the intervening region ${\cal R}$ of ${\mathscr I^+}$.
This computation serves as a check of the charge formula
(\ref{chargeans}).  We assume that the cuts ${\cal C}_1$ and ${\cal
  C}_2$ are in nonradiative vacuum regions of ${\mathscr I}^+$.

The flux in vacuum is proportional to the Bondi news tensor $N_{AB}$ and is
given by Eq.\ (82) of Wald and Zoupas \cite{Wald:1999wa}.
Translating this formula to Bondi coordinates
and adding the appropriate stress-energy flux gives for the total flux
\be
\int_{\cal R} d\Xi
=
- \int_{\cal R} \left(\frac 1{32\pi} N^{AB} \delta C_{AB}
+ \hat T_{ua}\xi^a \right) du\,d^2\Omega \, .
\label{check11}
\ee
Here the quantity $\delta C_{AB}$ is the change in $C_{AB}$ under the
BMS transformation ${\vec \xi}$, given by Eq.\ (\ref{deltaCAB}).

Since the conservation law (\ref{qformal1}) is linear in the generator
${\vec \xi}$, it is sufficient to verify the law separately for the
translation/supertranslation piece of ${\vec \xi}$, parameterized by $\alpha$, and
the remaining piece, parameterized by $Y^A$.  We first consider the
translation/supertranslation piece.  Using the expressions
(\ref{deltaCAB}) and (\ref{fdef}) for $\delta C_{AB}$ and
the formula (\ref{bmsinfinitesimal}) for the generator, specialized to
$Y^A=0$, and integrating by parts, we obtain
\begin{eqnarray}
\int_{\cal R} d\Xi
&=&
- \frac 1{32\pi} \int_{\mathscr I^+} \alpha \bigg(N^{AB}N_{AB} - 2 D_A D_B N^{AB}
\nonumber \\
&&+ 32\pi \hat T_{uu}\bigg) du\,d^2\Omega \, .
\end{eqnarray}
Using the evolution equation (\ref{dotm}) for the Bondi mass aspect and
assuming that the cuts ${\cal C}_1$ and ${\cal C}_2$ are of the form
$u=u_1$ and $u=u_2$ gives
\begin{eqnarray}
\int_{\cal R} d\Xi
&=&
 \frac 1{4\pi} \int_{{\cal C}_2} \alpha m \, d^2 \Omega -  \frac
 1{4\pi} \int_{{\cal C}_1} \alpha m \, d^2 \Omega,
\end{eqnarray}
which coincides with the required form (\ref{qformal1}) by Eq.\ (\ref{chargeans}).

Turn now to the Lorentz transformations parameterized by $Y^A$.
Inserting the expressions
(\ref{deltaCAB}) and (\ref{fdef}) for $\delta C_{AB}$ and
the formula (\ref{bmsinfinitesimal}) for the generator into Eq.\
(\ref{check11}), specializing
$\alpha=0$, and integrating by parts, we obtain
\be
\int_{\cal R} d \Xi = - \frac{1}{32 \pi} \int d^2 \Omega \int_{u_1}^{u_2} du \,
Y^A {\cal H}_A
\label{ch11}
\ee
where
\begin{eqnarray}
{\cal H}_A &=& -\frac{1}{2} u D_A (N_{BC} N^{BC}) + u D_A D_B D_C
N^{BC}\nonumber \\
&&+ \frac{1}{2} D_A (C_{BC} N^{BC}) + N^{BC} D_A C_{BC}
\nonumber \\
&& - 2 D_B(
N^{BC} C_{AC})
+ 32 \pi {\hat T}_{uA} - 16 \pi u D_A {\hat T}_{uu}. \ \ \
\end{eqnarray}

We next compute the change in the charge, given by the right hand side
of (\ref{qformal1}), to compare with (\ref{ch11}).
Differentiating the definition (\ref{hatNdef}) of ${\hat N}_A$ with
respect to $u$ and using the evolution equations
(\ref{dots}) together with (\ref{ccc}) gives
\begin{eqnarray}
&&\partial_u {\hat N}_A = - 8 \pi {\hat T}_{uA} + 4 \pi u D_A {\hat
  T}_{uu} + \frac{u}{8} D_A(N_{BC} N^{BC}) \nonumber \\
&&- \frac{3}{8} N_{AB} D_C C^{BC} + \frac{3}{8} C_{AB} D_C N^{BC}
+ \frac{1}{8} D_B C_{AC} N^{BC}
\nonumber \\
&&
 - \frac{1}{8} D_B N_{AC} C^{BC}
+ \frac{1}{4} D_B D_A D_C C^{BC} - \frac{1}{4} D_B D^B D^C C_{AC}
\nonumber \\
&&
- \frac{1}{4} u D_A D_B D_C N^{BC} - 2 \pi \partial_u {\hat T}_{rA}.
\label{hatNAdot}
\end{eqnarray}
Here we have also used the identities
\begin{subequations}
\begin{eqnarray}
D_A C_{BC} N^{BC} &=& D_B C_{CA} N^{BC} + N_{AB} D_C C^{BC}, \ \ \ \ \
\\
D_A N_{BC} C^{BC} &=& D_B N_{CA} C^{BC} + C_{AB} D_C N^{BC}, \ \ \ \ \
\end{eqnarray}
\end{subequations}
which can be verified by evaluating both sides in complex
stereographic coordinates $(z,{\bar z})$.

We now integrate Eq.\ (\ref{hatNAdot}) between $u_1$ and $u_2$,
use the expression (\ref{sr}) for the charge $Q$,
use the fact that ${\hat T}_{rA}$ vanishes at $u_1$ and $u_2$, and
compare with the flux (\ref{ch11}).  The result is
\be
\int_{\cal R} d \Xi = Q({\cal C}_2, {\vec \xi}) -Q({\cal C}_1, {\vec
  \xi}) + \Delta {\cal F},
\label{qformal2}
\ee
where the anomalous term is
\begin{eqnarray}
\Delta {\cal F} &=& \frac{1}{32 \pi} \int du  \int d^2 \Omega
\, Y^A \epsilon_{AB} \epsilon^{CD} D^B D_D D^E C_{CE}.\nonumber \\
\label{qformal3}
\end{eqnarray}
If we now decompose $C_{AB}$ into electric and magnetic parity pieces
according to Eq.\ (\ref{split}), we can rewrite this as
\begin{eqnarray}
\Delta {\cal F} &=& -\frac{1}{32 \pi} \int du  \int d^2 \Omega
\, Y^A \epsilon_{AB} D^B (D^2 + \frac{1}{2} D^4 ) \Psi. \nonumber \\
\label{anssx}
\end{eqnarray}
For BMS transformations, $Y^A$ is of the form
(\ref{lorentz}), where $\chi$ and $\kappa$ are purely $l=1$.
Integrating by parts, we see that the
expression (\ref{anssx}) vanishes, since $l=1$ harmonics are
annihilated by the operator $D^2 + 2$.
Hence we have verified the conservation law (\ref{qformal1}).

\section{Computation of superrotation charges in stationary regions in a more general class
  of Bondi frames}
\label{app:generalize}

The computations of the superrotation charges
in Sec.\ \ref{sec:charges1}
were specialized to Bondi frames
obeying the constraint (\ref{specialize}) that the angular momentum aspect be non-evolving.
While such frames always exist in stationary regions, they are not the most general frames.
We now extend those computations to remove this constraint, by using a different computational method.

As before, we restrict attention to a region of future null infinity
in which the spacetime is stationary, and in which the leading and subleading stress-energy tensor components vanish.
We start from the canonical Bondi frame (\ref{stat}),
and write the angular momentum aspect in that frame as
\be
N_A = \epsilon_{AB} D^B \Theta,
\label{canonical1}
\ee
where $\Theta$ is purely $l=1$ and independent of $u$ and encodes the intrinsic angular momentum.
We now perform a general linearized BMS transformation
parameterized by the vector field on future null infinity of the form [cf.\ Eq.\ (\ref{bmsinfinitesimal}) above, with some changes of notation]
\begin{align}
&{} \left[ - \frac{1}{2} \Phi + \frac{1}{2} u \lambda
\right] \partial_u + \left[ - \frac{1}{2} D^A \lambda + \epsilon^{AB} D_B \kappa \right] \partial_A.
\label{bmsinfinitesimal1}
\end{align}
Here $\Phi$, $\lambda$ and $\kappa$ are functions on the two-sphere which are independent of $u$.
The supertranslation piece $\Phi$ is arbitrary, while the boost piece $\lambda$ and the rotation piece $\kappa$ are $l=1$ harmonics.  We now combine this with the transformation laws (\ref{transforms}), the canonical form
(\ref{stat}) and (\ref{canonical1}) of the metric functions, the result (\ref{calD0}), and use $D_A D_B \lambda = - h_{AB} \lambda$ and similarly for $\kappa$.   Working to linear order in $\Phi$, $\lambda$ and $\kappa$, this yields for the metric functions in the new frame
\begin{subequations}
\label{stat1}
\begin{eqnarray}
\label{mstationary1}
m &=& m_0 + \frac{3}{2} m_0 \lambda, \\
\label{CABstationary1}
C_{AB} &=& D_A D_B \Phi - \frac{1}{2} h_{AB} D^2 \Phi, \\
N_A &=& \epsilon_{AB} D^B \left[ {\tilde \Theta} (1 + \lambda/2) \right] - \frac{3}{2} m_0 D_A \Phi
\nonumber \\ &&
 + \frac{3}{2} u m_0 D_A \lambda.
\end{eqnarray}
\end{subequations}
Here ${\tilde \Theta}$ is given by
\be
{\tilde \Theta} = \Theta + \epsilon^{AB} D_B \kappa D_A \Theta,
\ee
is purely $l=1$, and represents the intrinsic angular momentum in the rotated frame.
The results (\ref{stat1}) agree with the expressions (\ref{soln}) derived in Sec.\ \ref{sec:charges1} above,
except for the new terms involving the rotation $\kappa$ and the boost $\lambda$.

Next, we insert the results (\ref{stat1}) into the formula (\ref{chargeans}) for the BMS charge,
specialized to $\alpha =0$, neglecting terms quadratic in $C_{AB}$ as before.  This gives
\be
Q = \frac{1}{8 \pi} \int d^2 \Omega Y^A {\hat N}_A,
\label{chargeans2}
\ee
where
\be
{\hat N}_A =
 \epsilon_{AB} D^B \left[ {\tilde \Theta} (1 + \lambda/2) \right] - \frac{3}{2} m_0 D_A \Phi.
\label{fff}
\ee
Comparing this with Eq.\ (\ref{soln}) we see that there is an extra
contribution to the magnetic parity piece of the superrotation charges
$Q[l_m]$ for $|m| > 1$ associated with the boost piece $\lambda$ of the BMS transformation.  It gives a non-zero contribution
only for $|m| =  2$.

\section{Magnetic parity piece of shear vanishes in
  stationary vacuum regions}
\label{app:magzero}

In this appendix we show that the magnetic parity piece of the shear
tensor $C_{AB}$ vanishes in stationary vacuum regions of future null infinity
${\mathscr I}^+$, in arbitrary Bondi frames.
Closely related results
have been derived by Winicour
\cite{2014CQGra..31t5003W,2016CQGra..33q5006M},
and by Bieri and Garfinkle \cite{Bieri:2013ada} in linearized gravity.

In the canonical Bondi frame discussed in Sec.\ \ref{sec:stationary} above, the
metric functions take the simple form (\ref{stat}), and in particular $C_{AB}=0$.
Consider now making a transformation to an arbitrary Bondi frame,
using the general nonlinear BMS transformation discussed in Appendix
\ref{app:finite}.
We can decompose such a transformation into a rotation, followed by a
boost, followed by a supertranslation.  The rotation and boost maintain
$C_{AB}=0$, by Eq.\ (\ref{dd}).  Finally, under the supertranslation
the shear tensor undergoes the transformation (\ref{st11}), which
generates only an electric parity piece of $C_{AB}$.

\bibliography{infoloss}

\end{document}